\documentclass{article}
\usepackage{proceed2e}

\usepackage{geometry}
\usepackage{times}
\usepackage[numbers,sort&compress]{natbib}
\usepackage{amsmath}
\usepackage{amssymb}
\usepackage{graphicx}
\usepackage{caption}
\usepackage{subcaption}
\graphicspath{{./}{./figures/}{../figures/}}
\DeclareGraphicsExtensions{.pdf,.png,.jpg}
\usepackage{multirow}
\usepackage{booktabs}
\usepackage{tabularx}
\usepackage{siunitx}
\usepackage{url}
\usepackage{hyperref}
\hypersetup{
     colorlinks   = true,
     allcolors    = NavyBlue,
}
\usepackage[capitalize]{cleveref}

\newcommand{\subfigimg}[4][,]{%
  \setbox1=\hbox{\includegraphics[#1]{#3}}
  \leavevmode\usebox1\newline
  \makebox[\linewidth]{\raisebox{0\ht1}{\small\textbf{#2} #4}}
}

\usepackage[dvipsnames]{xcolor}

\newcommand{\vn}[1]{\mathrm{#1}} 

\title{Generating Realistic X-ray Scattering Images Using Stable Diffusion and Human-in-the-loop Annotations}

\author{ 
Zhuowen Zhao$^{1,2}$ 
\And
Xiaoya Chong$^1$  
\And 
Tanny Chavez$^1$
\And
Alexander Hexemer$^1$
\AND 
\textnormal{$^1$Advanced Light Source (ALS), Lawrence Berkeley National Laboratory, Berkeley, CA 94720} \\
\textnormal{$^2$Chan Zuckerberg Institute for Advanced Biological Imaging (CZ Imaging Institute), Redwood City, CA 94065} \\
\textnormal{kevin.zhao@czii.org, xchong@lbl.gov, tanchavez@lbl.gov, ahexemer@lbl.gov}
}

\begin{document}

\maketitle

\begin{abstract}
We fine-tuned a foundational stable diffusion model using X-ray scattering images and their corresponding descriptions to generate new scientific images from given prompts. 
However, some of the generated images exhibit significant unrealistic artifacts, commonly known as ``hallucinations''. 
To address this issue, we trained various computer vision models on a dataset composed of \SI[round-precision=0]{60}{\percent} human-approved generated images and \SI[round-precision=0]{40}{\percent} experimental images to detect unrealistic images. 
The classified images were then reviewed and corrected by human experts, and subsequently used to further refine the classifiers in next rounds of training and inference.
Our evaluations demonstrate the feasibility of generating high-fidelity, domain-specific images using a fine-tuned diffusion model.
We anticipate that generative AI will play a crucial role in enhancing data augmentation and driving the development of digital twins in scientific research facilities.
\end{abstract}

\section{Introduction}
Artificial intelligence (AI) is becoming an increasingly important tool for tackling many complex challenges in today's scientific research.
It is used for enhancing the precision and automation of high-throughput experiments at user facilities \cite{CORREABAENA20181410, mlexchange2022, autonomous_microscope2023}, improving the analysis of large and multimodal high-dimensional datasets \cite{ZHAO202071, Zubatiuk_etal2021, Amal_etal2022, Ektefaie_etal2023}, and enabling the accurate and rapid detection of fine and heterogeneous particles and features in biomedical imaging \cite{CryoDRGN, tomotwin, bioengineering10121435, tibbers_etal2023}.
Training complex AI models and systems for these applications often requires large curated datasets.
However, establishing high-quality training datasets through experiments can be very costly. 
Thus, developing ultra-realistic data augmentation methods is crucial to overcome this challenge.

X-ray scattering images are integral to synchrotron research, providing valuable insights into the structural properties of materials. 
Several approaches have been explored to synthesize X-ray images, including physics-based simulations and ray tracing techniques that aim to approximate real-world conditions \cite{khaira2017derivation}. 
While these methods are effective in certain contexts, they often struggle to capture the complex interactions and subtle nuances present in actual X-ray data. 
In recent years, generative adversarial networks (GANs) have been utilized to generate synthetic X-ray images. 
This deep learning-based approach has shown promise in producing more accurate and visually appealing X-ray images. For example, Guo et al. \cite{guo2022physics} employed GANs to learn the distribution of X-ray tomography data, generating synthetic images with enhanced realism. 
However, GANs are notoriously difficult to train, requiring extensive computational resources and expertise \cite{pmlr-v80-mescheder18a, 10565846, NEURIPS2020_3eb46aa5}.

Recent advances in generative diffusion models have demonstrated their ability \cite{bai2022traininghelpfulharmlessassistant} to interpret inputs and generate outputs in different modalities, such as text (text-to-text) \cite{brown2020languagemodelsfewshotlearners}, images (text-to-image) \cite{rombach2022high}, and videos (text-to-video) \cite{liu2024sorareviewbackgroundtechnology, maaz2024videochatgptdetailedvideounderstanding}. 
While these generative techniques have greatly expanded the possibilities for artistic and visual creativity, the latest models have also shown potential in capturing visuality following physical laws \cite{ho2020denoisingdiffusionprobabilisticmodels, song2022denoisingdiffusionimplicitmodels, rombach2022high, podell2023sdxl, videoworldsimulators2024}.
This suggests that generative diffusion models could be viable tools for augmenting image datasets to meet the demands of training complex AI systems in scientific applications. 
Although there have been reports of diffusion models being used to generate medical X-ray images, their application to X-ray scattering images in synchrotron settings has not been explored.
For instance, Chambon et al. \cite{chambon2022adapting} adapted the pretrained vision-language foundational model stable diffusion to generate domain-specific images for chest X-ray images. 
Most recently, Hashmi et al. \cite{hashmi2024xreal} used a diffusion model to generate chest X-ray images with spatial control over anatomy and pathology. 
Liang et al. \cite{liang2024covid} fine-tuned a stable diffusion model to synthesize high-resolution chest X-ray images (\(512\times512\) pixels) depicting bilateral lung edema caused by COVID-19 pneumonia using a class-specific prior preservation strategy.
 
In this work, we propose a pipeline that leverages a foundational stable diffusion model and various computer vision models to create realistic X-ray scattering images. 
Additionally, we developed a continuous training framework that incorporates human annotations into the process. 
It is worth mentioning that this paper does not focus on the most recent generative diffusion model implementations. 
Our primary objectives are to access the quality of the generated images and to investigate methods for enhancing them within the computational resources available to a scientific research group.

\section{Methodology}
\begin{figure*}[h!]
\centering
\includegraphics[width=\linewidth]{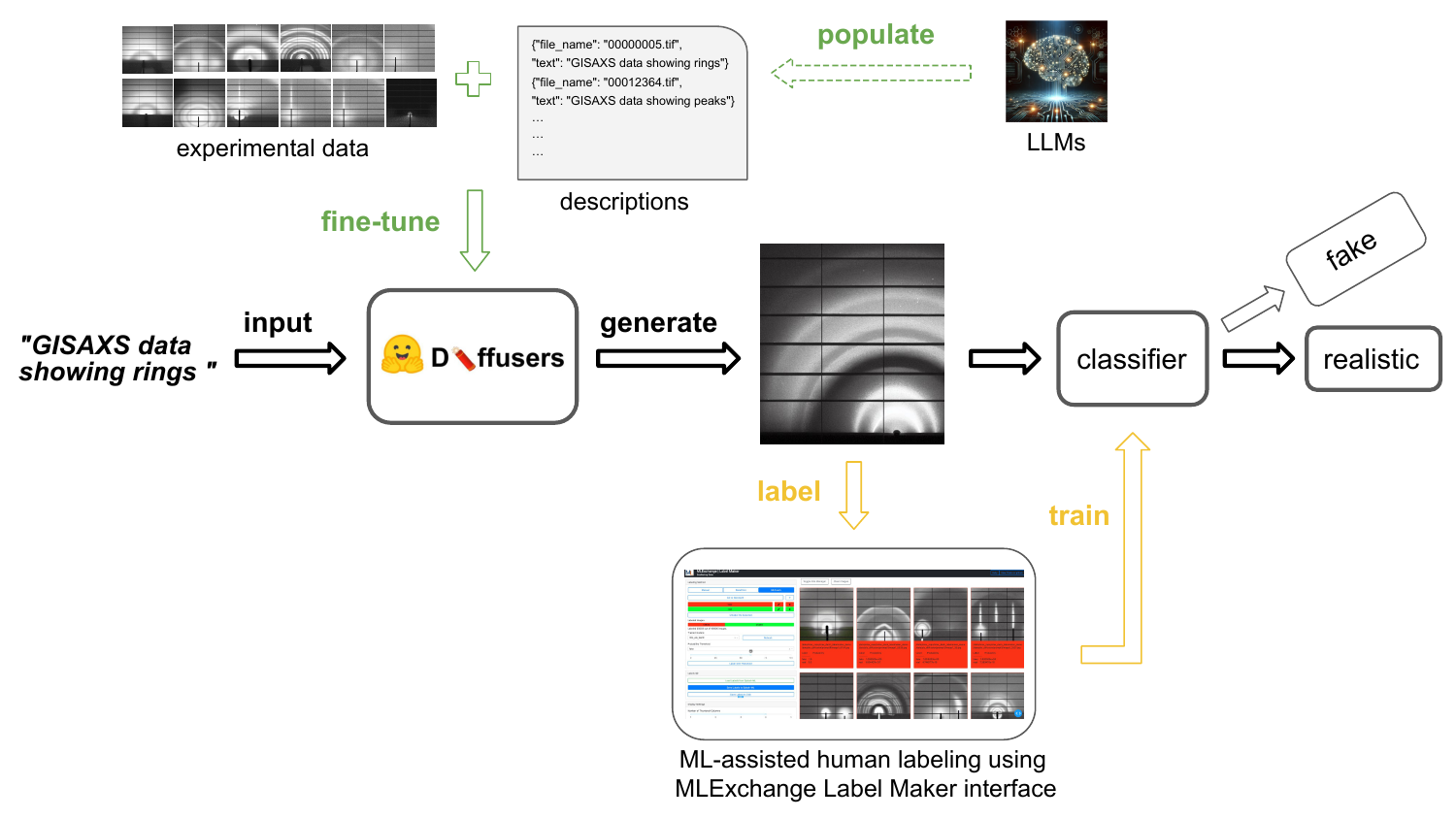}
\caption{The schematics showing the pipeline to create realistic X-ray scattering images using generative AI and human annotations. The bold black arrows demonstrate the process to generate data from an input prompt.}
\label{fig: architecture}
\end{figure*}

\Cref{fig: architecture} illustrates the framework for generating and refining X-ray scattering images using a foundational diffusion model and continuous human annotations. 
The pipeline comprises the following steps: (a) we fine-tuned a foundational diffusion model with the curated X-ray scattering images and their text descriptions; (b) the generated images are labeled as ``realistic'' or ``fake'' by domain experts; (c) a ResNet-50 model was trained with ImageNet weights based upon the human labels to classify the unseen generated images; (d) we iterated (b) and (c) to increase the number of labeled images. For each iteration, we retrained an assortment of foundational computer vision models (Vision Transformer, ResNet-50, VGG, etc.) using ensemble classification strategies to maximize the detection of realistic X-ray scattering images.%
\footnote{We chose to not include the diffusion model in the iterative re-training with human annotations based on two considerations: 1. the diffusion model can be replaced with a new foundational model without the need to retrain the whole pipeline; 2. the pipeline avoids the GAN-style architecture that is known for unstable training \cite{pmlr-v80-mescheder18a, 10565846, NEURIPS2020_3eb46aa5}.}

\subsection{Stable diffusion model}
Diffusion models represent probabilistic frameworks aimed at learning a data distribution $p(x)$ through the gradual denoising of a normally distributed variable. This process mirrors the reverse learning of a fixed-length Markov Chain. In the context of image synthesis, these models function as a series of denoising autoencoders $\epsilon_{\theta}(x_{t}, t); t = 1...T$, each tasked with predicting a cleaner version of their input $x_{t}$, where $x_{t}$ represents a noisy variant of the original input $x$. The objective associated with this process can be simplified as $L_{DM} = E_{x,\epsilon \sim N(0,1),t}\left [ \left \| \epsilon -\epsilon _{\theta }(x_{t},t) \right \|_{2}^{2} \right ]$, with $t$ uniformly sampled from $\left \{ 1, . . . , T  \right \}$.

To facilitate training diffusion models on constrained computational resources while preserving their quality and adaptability, stable diffusion model utilizes them within the latent space of potent pretrained autoencoders. Specifically, given an image $x \in \mathbb{R}^{H \times W \times 3}$, the encoder $\varepsilon$ encodes $x$ into a latent representation $z = \varepsilon(x)$, and the decoder $D$ reconstructs the image from this latent space, yielding $\tilde{x} = D(z) = D(\varepsilon(x))$, where $z \in \mathbb{R}^{h \times w \times c}$. 
With a trained $\varepsilon$ and $D$, we gain access to an efficient, low-dimensional latent space wherein high-frequency, imperceptible details are abstracted. The generative modeling of latent representations can be expressed as $L_{DM} = E_{x,\epsilon \sim N(0,1),t}\left [ \left \| \epsilon -\epsilon _{\theta }(z_{t},t) \right \|_{2}^{2} \right ]$.

The model's neural backbone, $\epsilon_{\theta}(\cdot, t)$, is implemented as a time-conditional U-Net. Given that the forward process remains constant, $z_{t}$ can be efficiently derived from $\varepsilon$ during training, while samples from $p(z)$ can be decoded into image space with just one pass through $D$.

To preprocess $y$ originating from prompts, a language encoder $\tau_{\theta}$ is used to transform $y$ into an intermediate representation $\tau_{\theta}(y) \in \mathbb{R}^{M \times d_{\tau}}$. Subsequently, this representation is integrated into the intermediate layers of the U-Net using a cross-attention layer that implements $Attention(Q,K,V) = \text{softmax}(\frac{QK^{T}}{\sqrt{d}})$. The conditional learning of LDM is then facilitated through $L_{DM} = E_{x,\epsilon \sim N(0,1),t}\left [ \left \| \epsilon -\epsilon _{\theta }(z_{t},t,\tau_{\theta }(y)) \right \|_{2}^{2} \right ]$.

In this work, we utilized the Hugging Face implementation of a stable diffusion model \cite{diffusers}, known as Diffusers (SD 1.5). 
This model was pre-trained on a subset of the LAION-5B image database \cite{rombach2022high}. 
It incorporates CLIP ViT-L as the text encoder, which has 123 million parameters, conditioning the model on textual prompts alongside a U-Net.  
The specifications of the Diffusers are detailed in \cref{tab:sd}.

\begin{table}[h!]
	\caption{Model statistics of SD 1.5}
	\label{tab:sd}
	\centering
	\begin{tabular}{cc} 
	\toprule
        U-Net parameters  & 860 million          \\ 
        transformer blocks & {[}1,1,1,1{]} \\ 
        channel multiplier     & {[}1,2,4,4{]} \\ 
        text encoder       & CLIP ViT-L/14    \\ 
        context dimension       & 768           \\ 
        	\bottomrule
	\end{tabular}
\end{table}

\subsection{Training dataset for Diffusers}
\label{sec: training datasets}

\begin{figure*}[h!]
\centering
\begin{tabular}{ccc}
\includegraphics[width=0.15\linewidth]{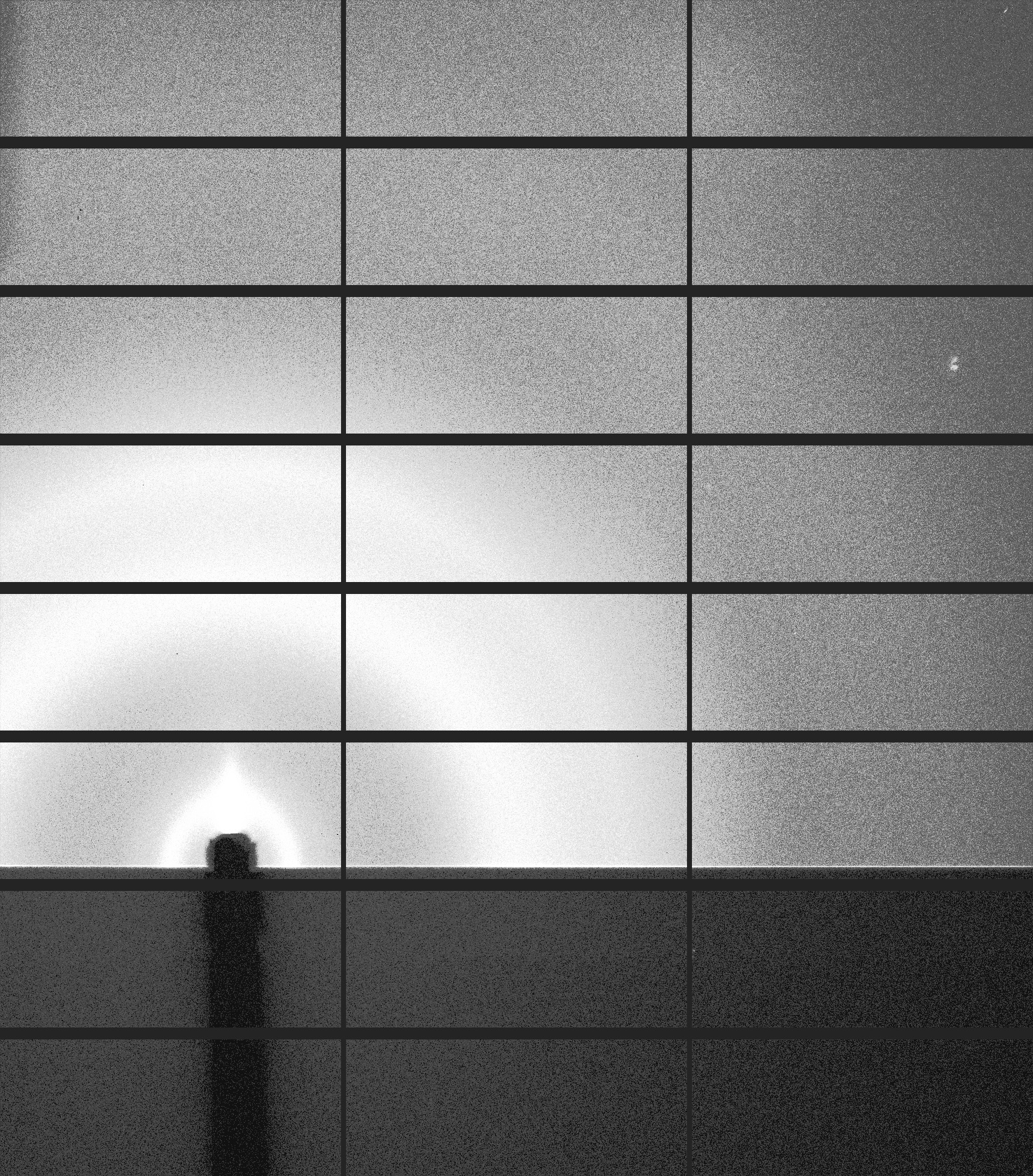} 
\includegraphics[width=0.15\linewidth]{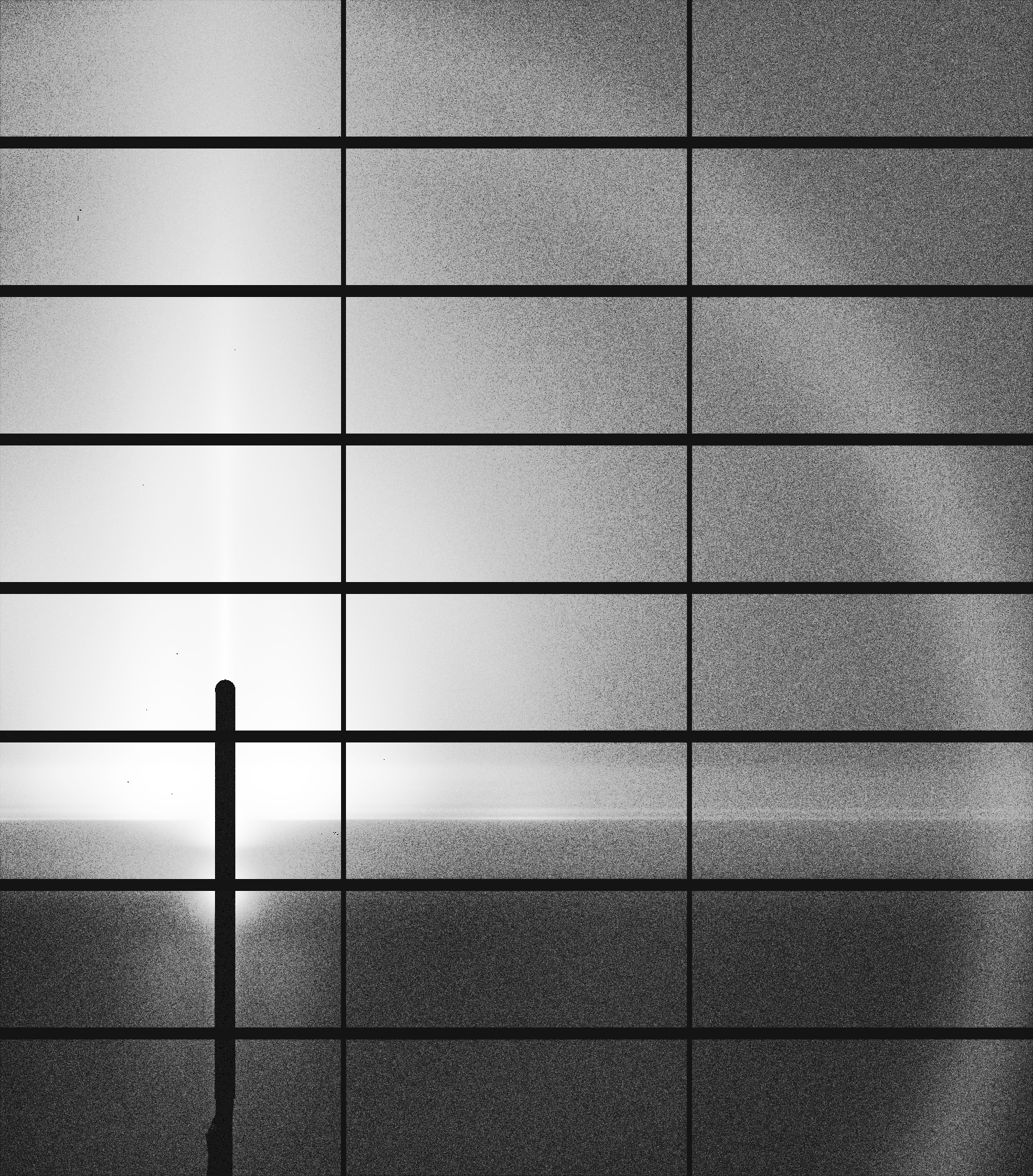} 
&
\includegraphics[width=0.15\linewidth]{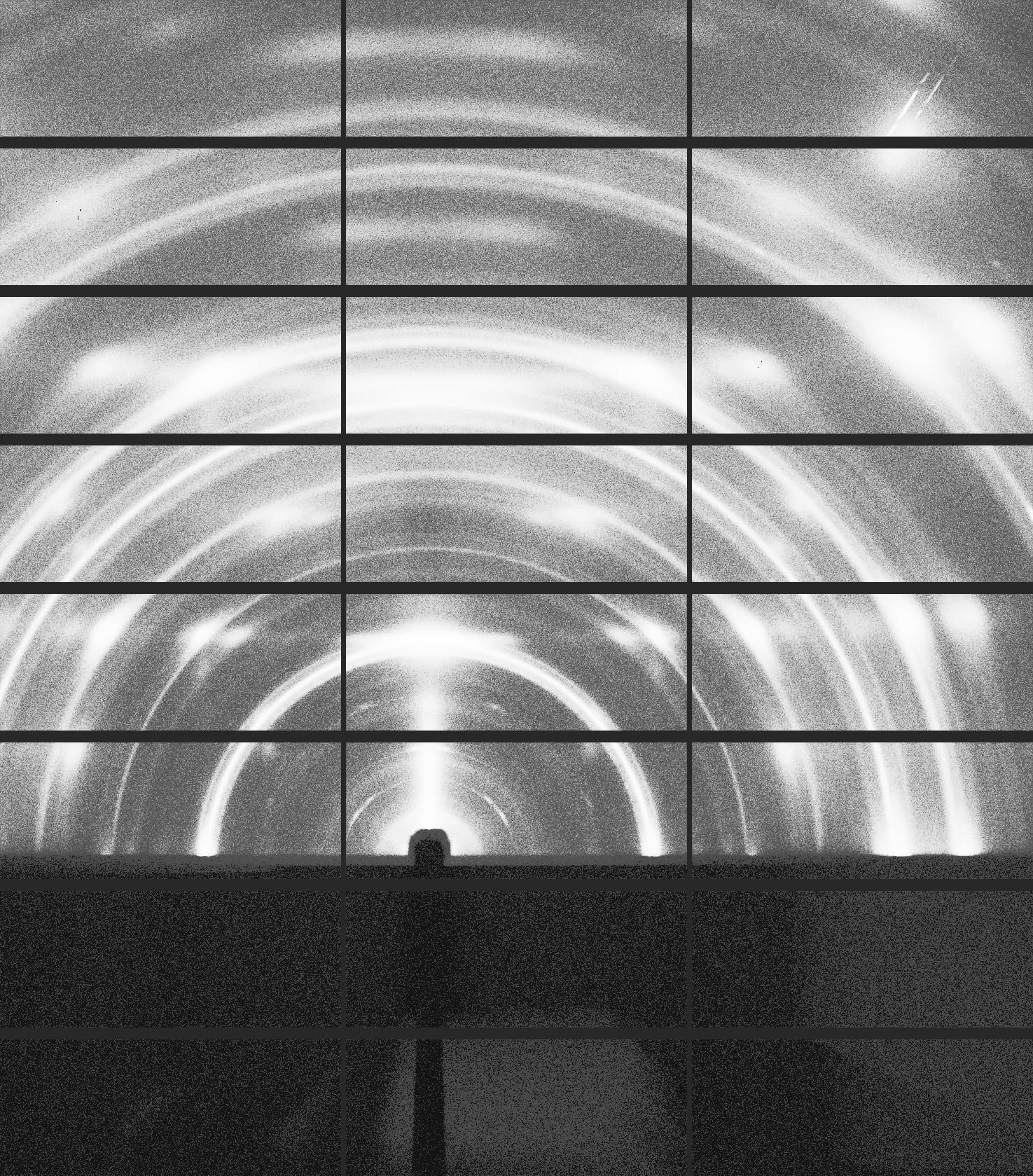}
\includegraphics[width=0.15\linewidth]{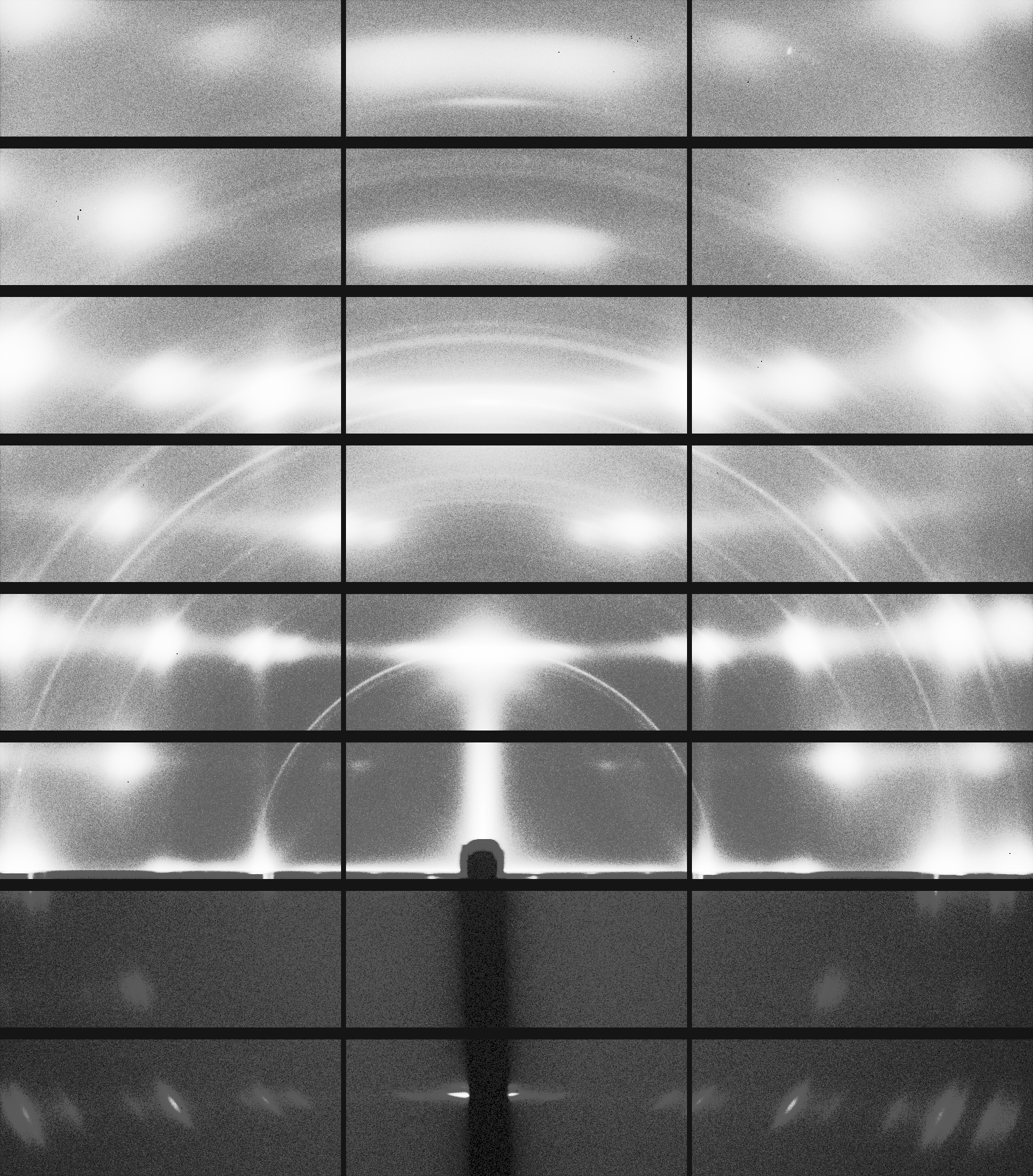}  
&
\includegraphics[width=0.15\linewidth]{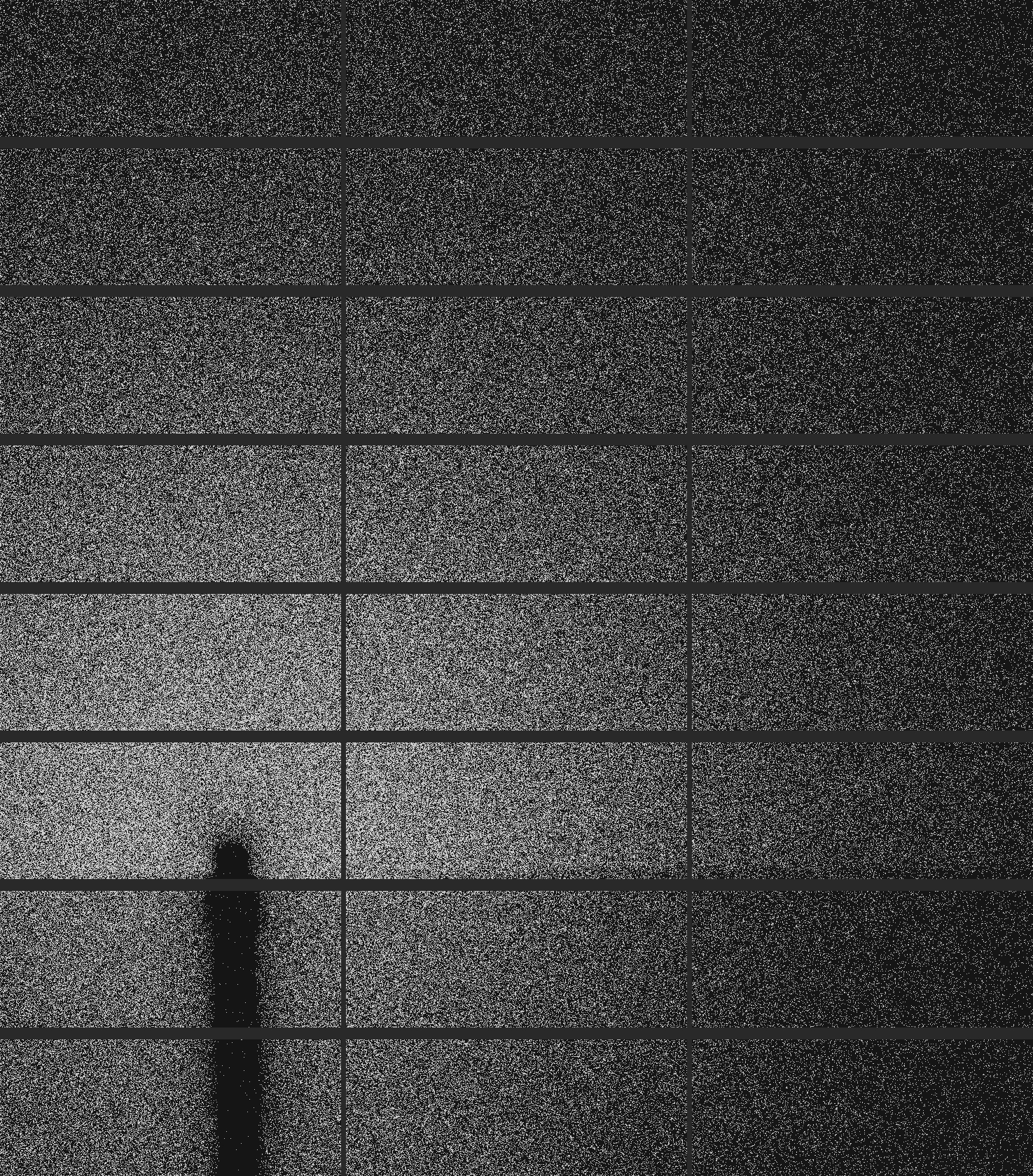} 
\includegraphics[width=0.15\linewidth]{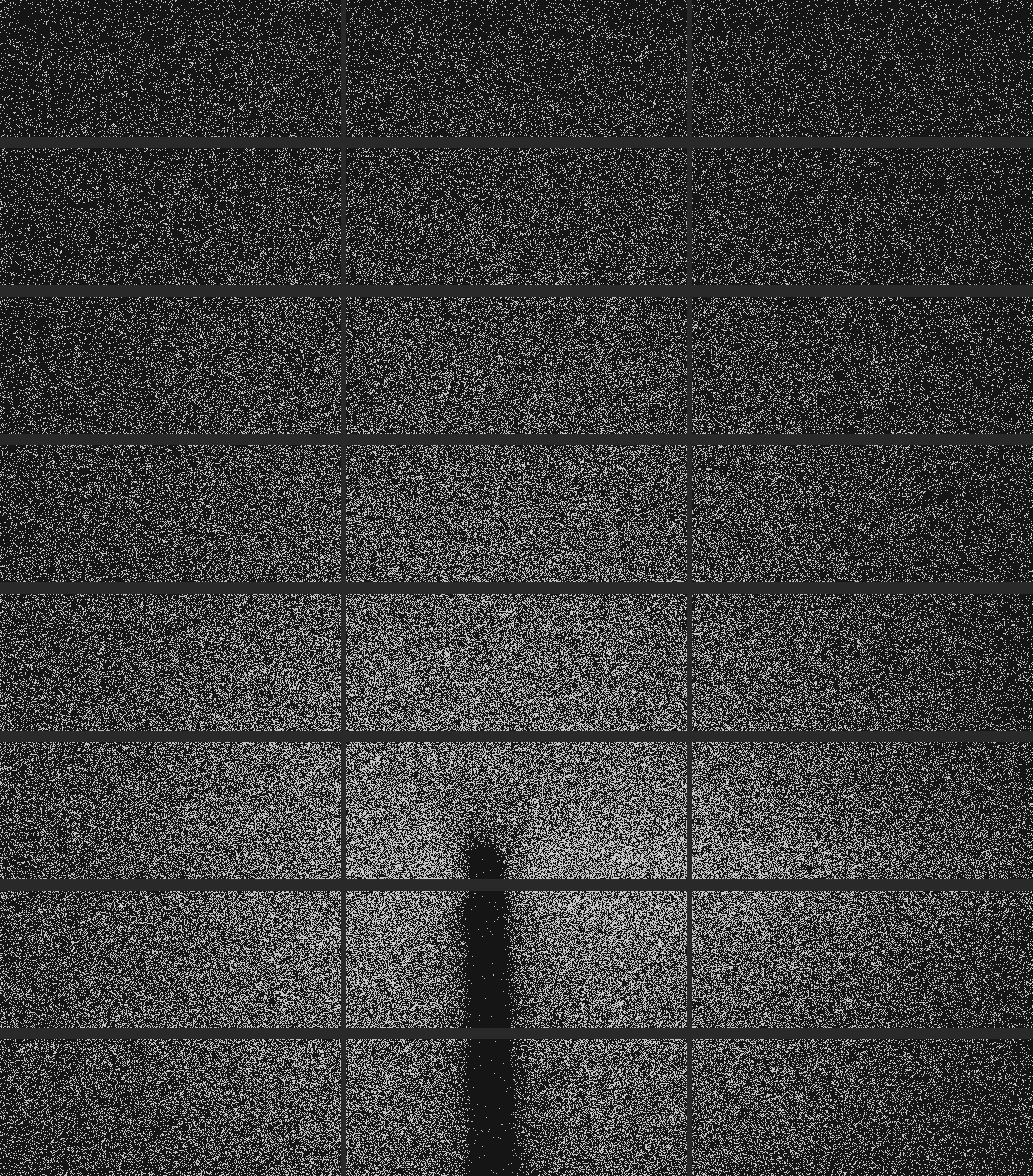}  
\\
\includegraphics[width=0.15\linewidth]{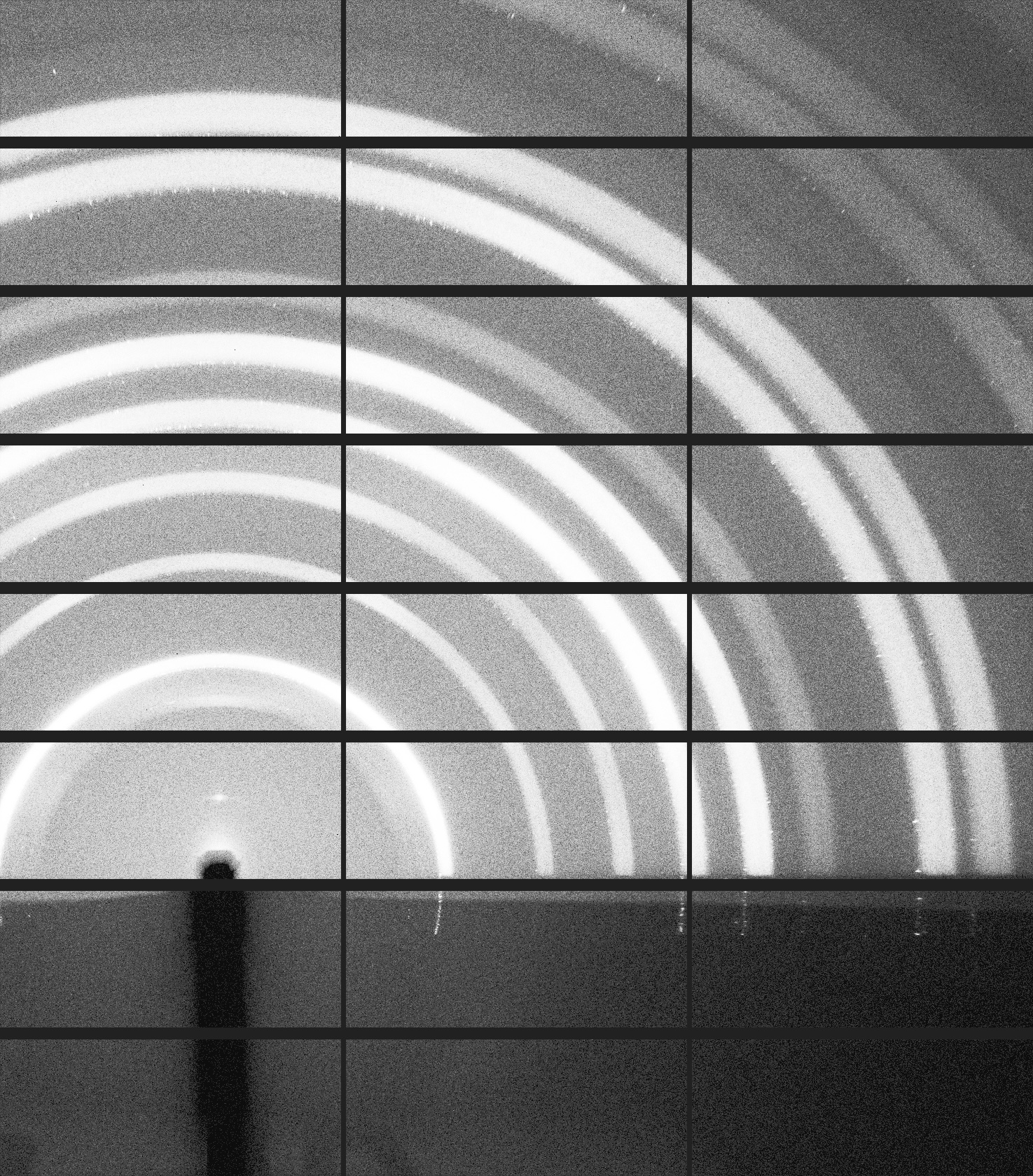} 
\includegraphics[width=0.15\linewidth]{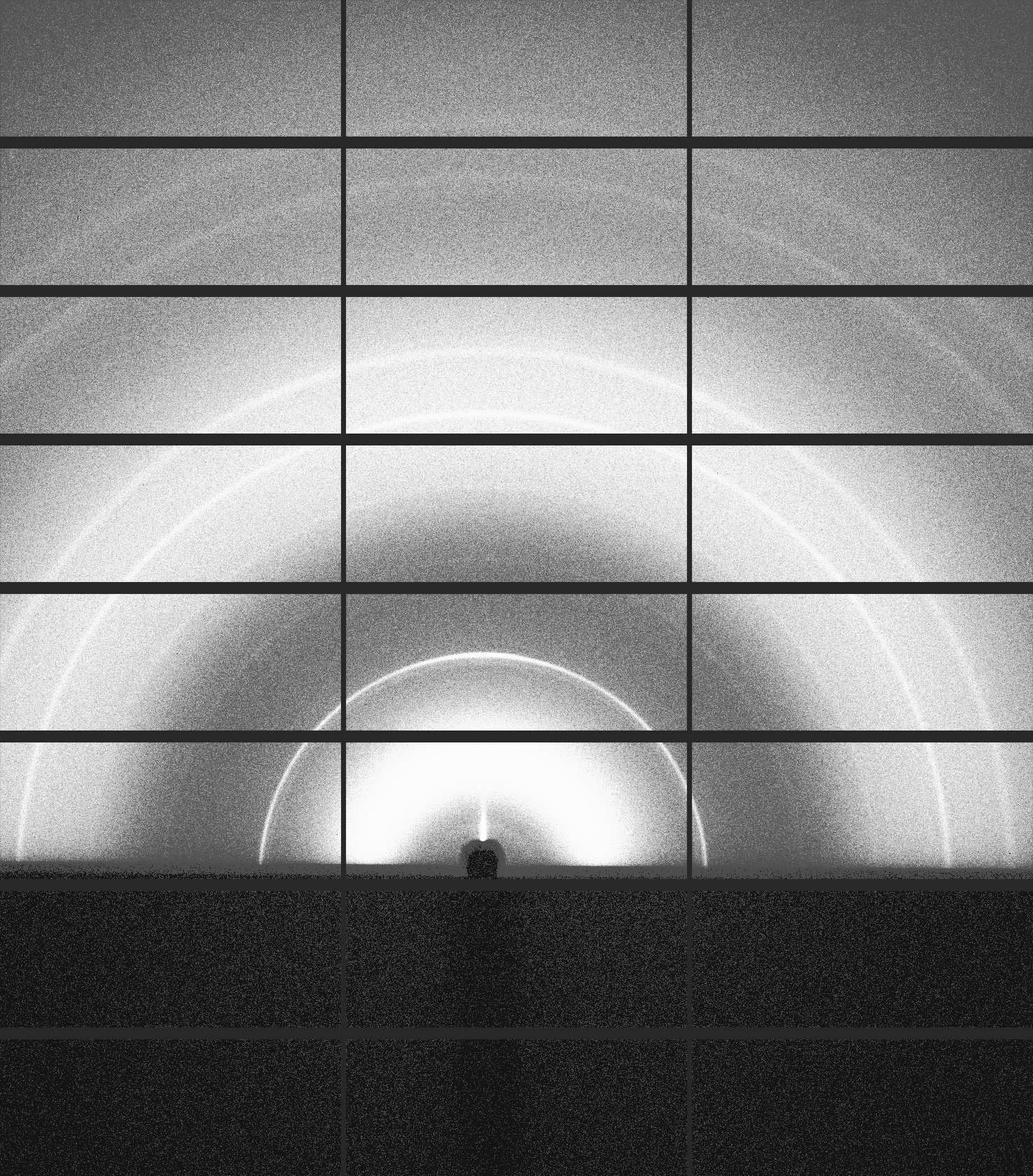} 
&
\includegraphics[width=0.15\linewidth]{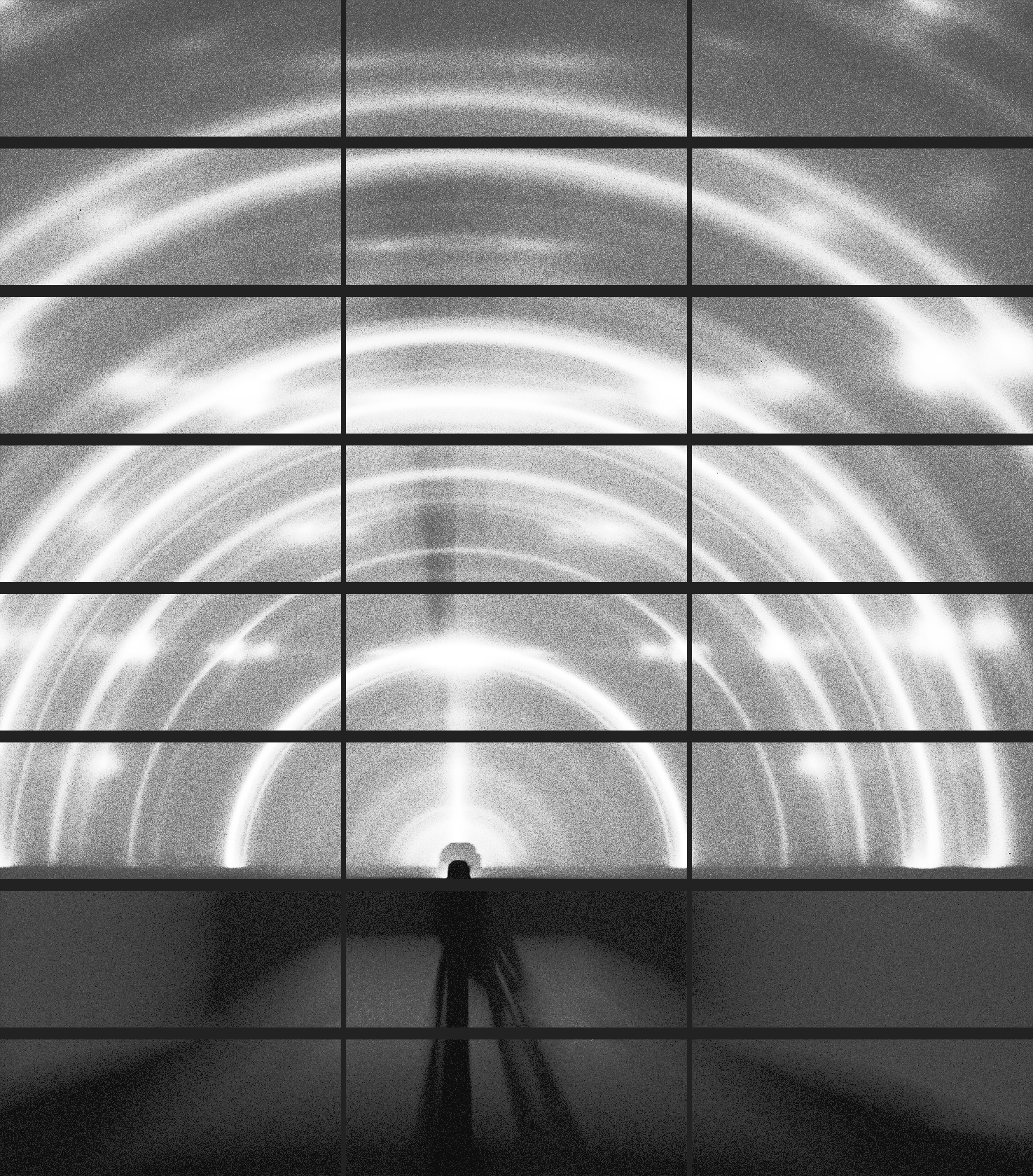}
\includegraphics[width=0.15\linewidth]{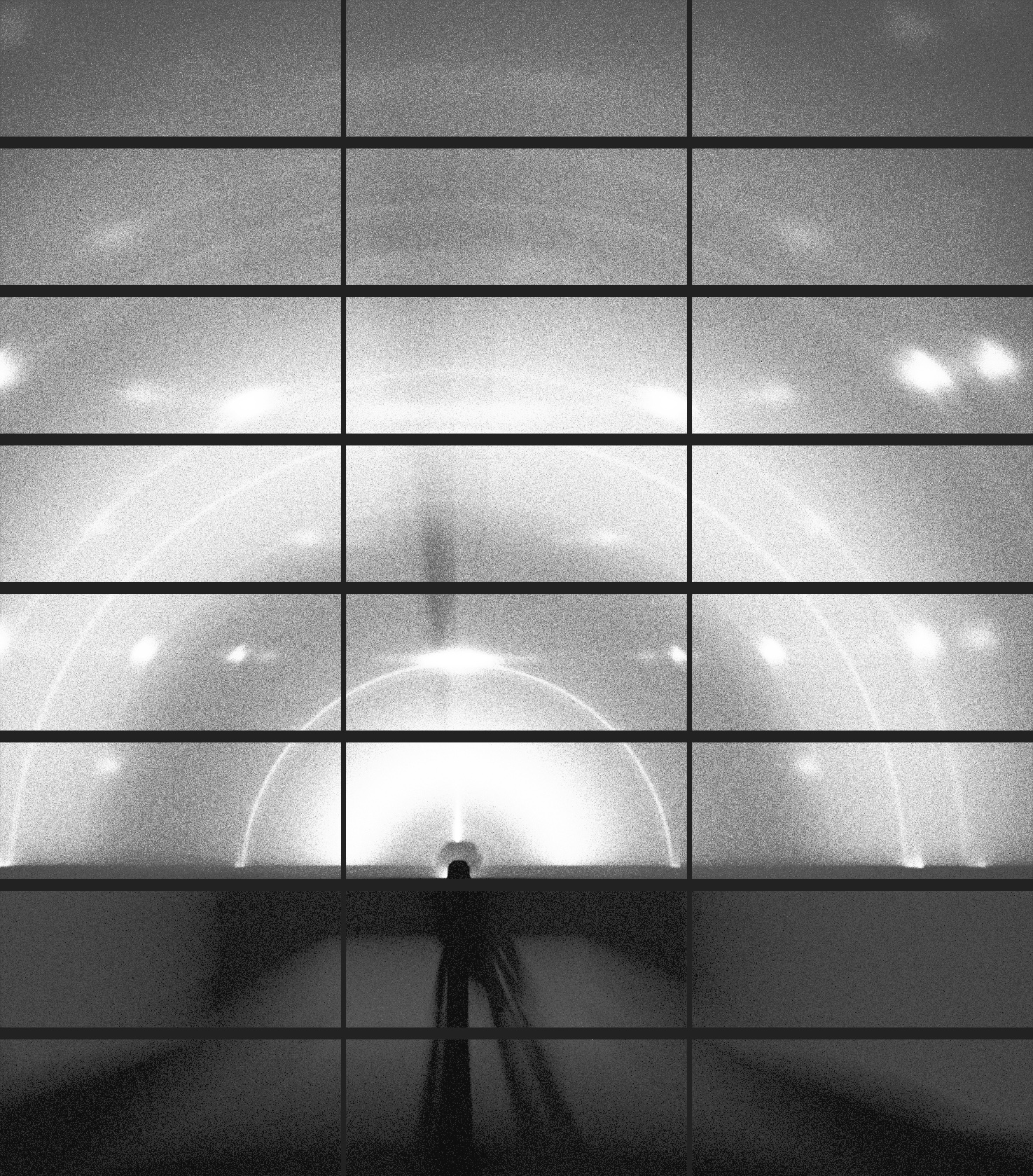}  
&
\includegraphics[width=0.15\linewidth]{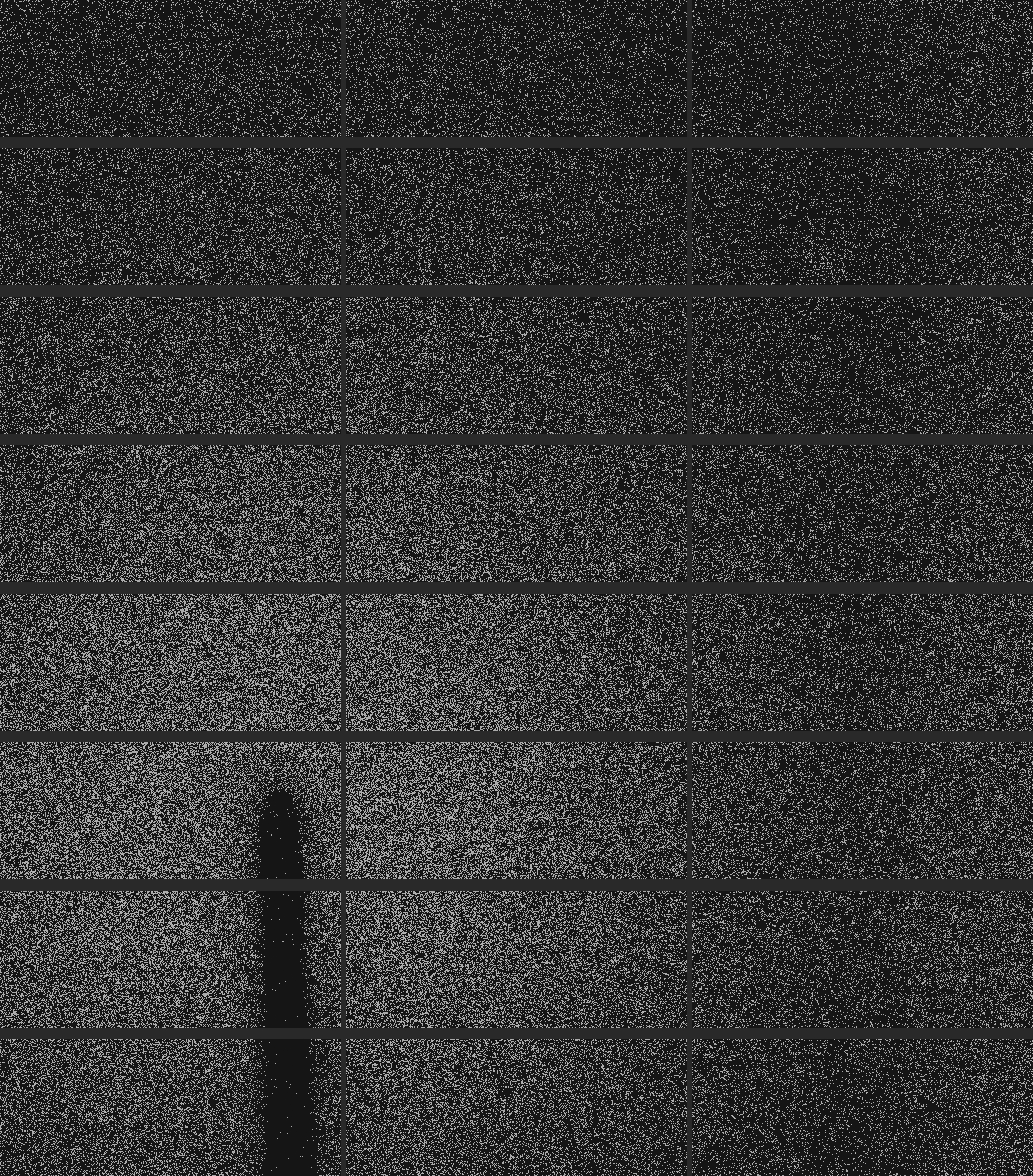} 
\includegraphics[width=0.15\linewidth]{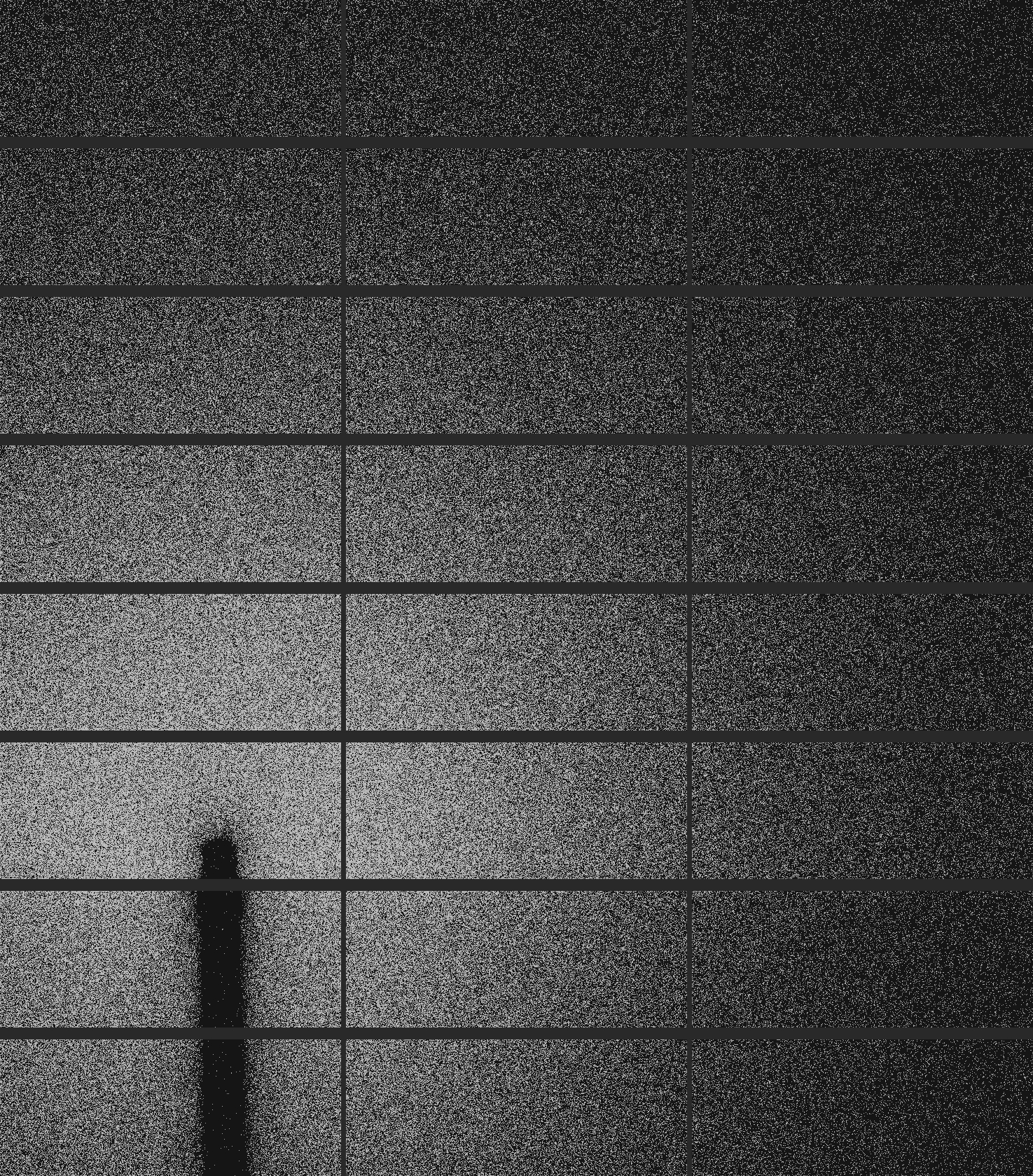} 
 \\
rings & peaks & background  
\end{tabular}
\caption{Examples of experimental images used for fine-tuning the Diffusers model.}
\label{fig: exp}
\end{figure*}

\num[round-precision=0]{300} X-ray scattering images were selected to fine-tune the Diffusers model.\footnote{The training dataset includes images from grazing-incidence small-angle X-ray scattering (GISAXS), small-angle X-ray scattering (SAXS), and wide-angle X-ray scattering (WAXS), all collected at beamline 733 of the Advanced Light Source, Berkeley Lab.}
These images encompass three main patterns: background (empty frame with a beamstop), peaks, and rings, with \num[round-precision=0]{100} images per pattern (see \cref{fig: exp}).
A metadata file, containing the path of each image along with its corresponding textual description, is used by the Diffusers model to load text-image pairs in batches during training.

It is important to note that the majority of the experimental scattering images are rings, with peak images being very rarely observed (only a few per \num[round-precision=0]{1000} experimental images) and exhibiting significant similarity to each other. 
We identified only a handful of background images in the entire experimental dataset of \num[round-precision=0]{400000} images. Consequently, we augmented the background images to increase the size of the training dataset to \num[round-precision=0]{100} images.
As a result, the selected training dataset for rings exhibits higher diversity compared to the datasets for peaks and background images.

\subsection{Classification with continuous training and human-in-the-loop annotations}
\label{sec: enssemble classification}

\begin{figure*}[h!]
\centering
\begin{tabular}{cc}
\includegraphics[width=0.45\linewidth]{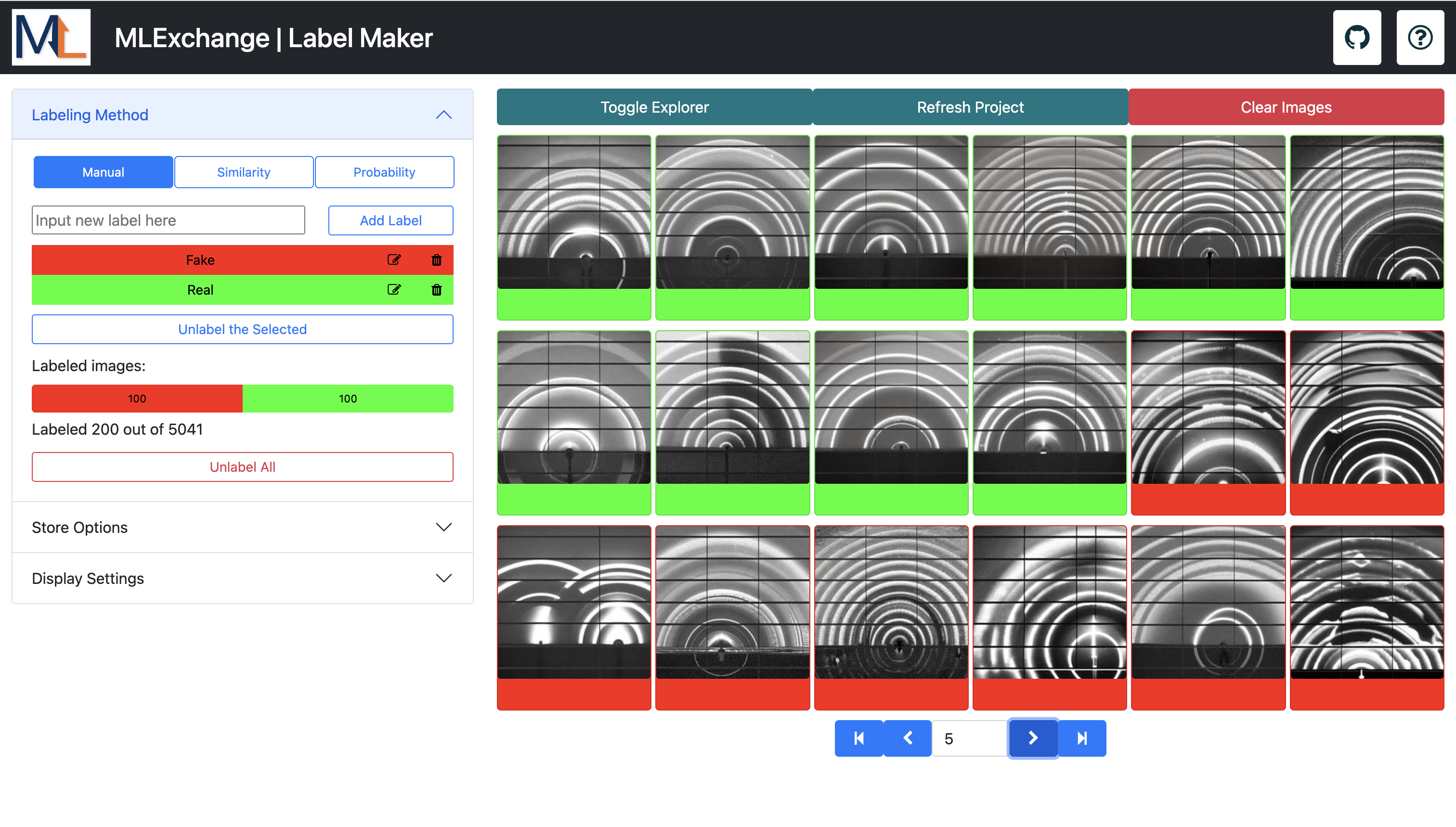} 
&
\includegraphics[width=0.45\linewidth]{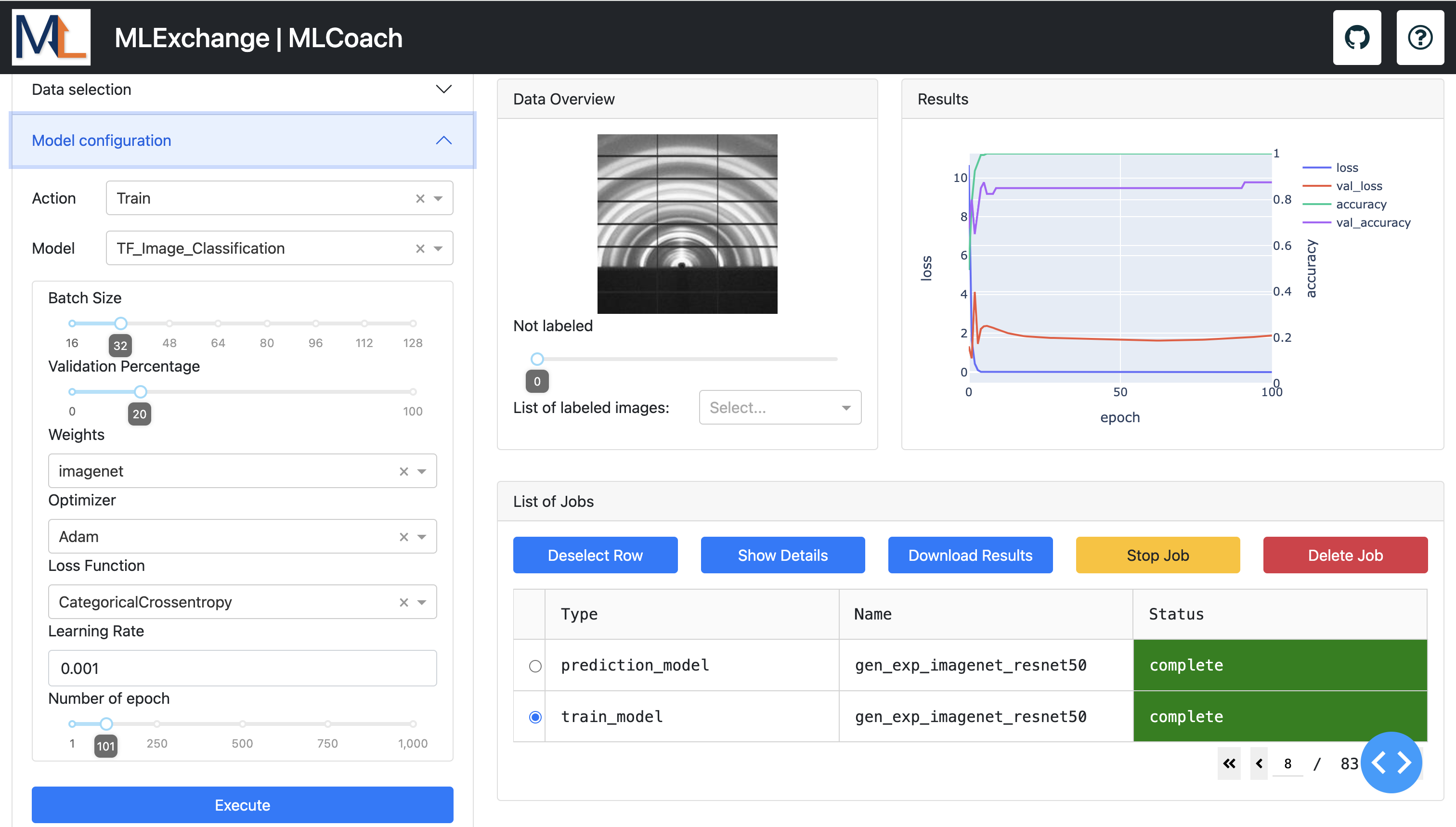} \\
Label Maker & MLCoach 
\end{tabular}
\caption{Label Maker and MLCoach web interface}
\label{fig: mlex label}
\end{figure*}

\begin{figure*}
\centering
\includegraphics[width=\linewidth]{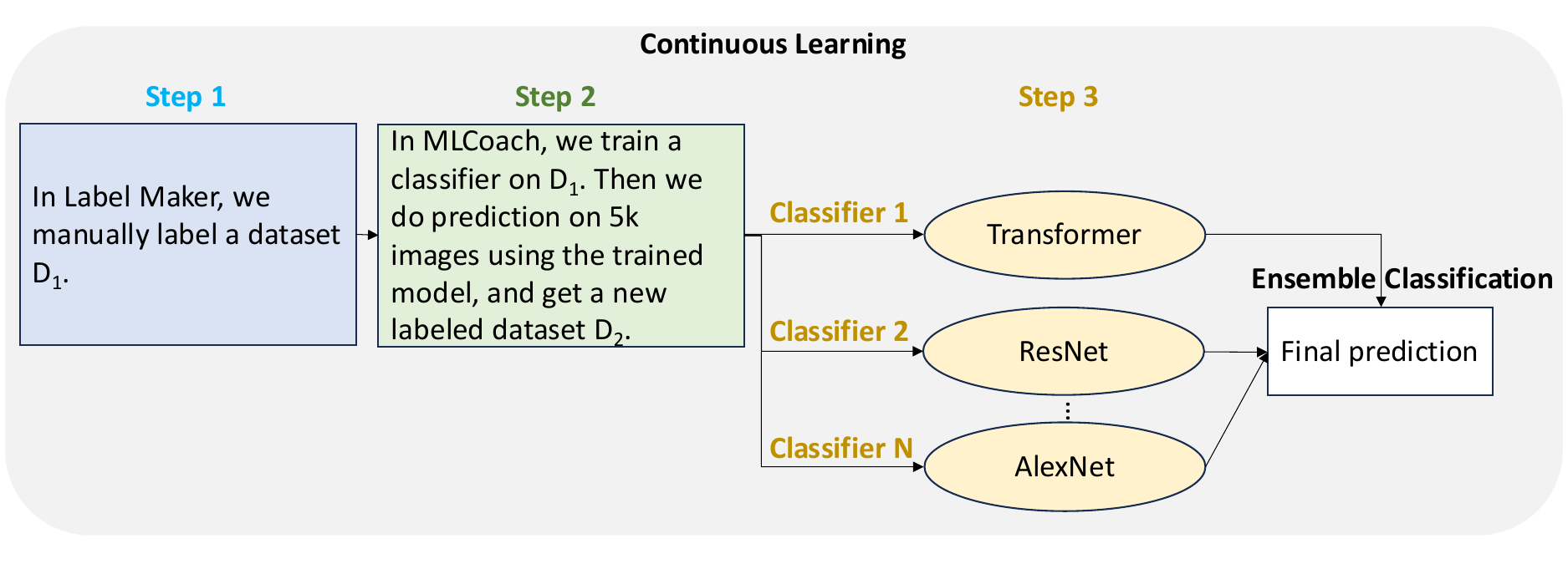}
\caption{Continuous training with human-in-the-loop annotations and ensemble classification}
\label{fig: RLHF classification}
\end{figure*}

The fine-tuned Diffusers model generated \num[round-precision=0]{40000} images in about \num[round-precision=0]{4} hours.
However, some of the generated images exhibit unrealistic artifacts, a phenomenon widely observed in generative models referred to as hallucinations \cite{aithal2024understandinghallucinationsdiffusionmodels, weng2024hallucination}.
These artifacts may originate from various sources. 
For instance, Aithal et.al. \cite{aithal2024understandinghallucinationsdiffusionmodels} recently reported that hallucinations can result from smooth interpolations of distributions across ``nearby data modes'' in the training dataset when generating images outside the original training dataset distribution. Additionally, since both the diffusion model and text encoder were pre-trained with non-scientific training datasets, the Diffusers model can generate features from other domains.  
Thus, we trained deep learning models as classifiers for selecting the images that are realistic.
Since the number of labeled images is crucial for the training, we developed an iterative strategy to establish human annotations with the help of ResNet-50 model, leveraging our in-house interactive labeling tools, Label Maker and MLCoach \cite{zhao2022mlexchange}, as depicted in \cref{fig: mlex label}. 
The detailed steps of our strategy are illustrated in \Cref{fig: RLHF classification} and are described below.

\paragraph{Step 1: creating a training and validation dataset with human annotations} In Label Maker, we manually labeled \num[round-precision=0]{60} realistic and \num[round-precision=0]{100} fake images from a set of \num[round-precision=0]{1000} generated images (rings or peaks respectively). 
The metrics for determining a realistic image is discussed in \cref{sec: generated images}.
Since a hybrid dataset ($\text{D}_1$) using both experimental and generated images (maintaining a 4 to 6 ratio in all iterations) typically yields better training results, we augmented the realistic subset by adding \num[round-precision=0]{40} experimental images, making its size equal to that of the fake subset. 
In addition, we created a distinct validation dataset in the same way.

\paragraph{Step 2: creating a larger training dataset with ResNet-50 and human annotations} We trained a ResNet-50 model using ImageNet weights and $\text{D}_1$ in MLCoach. 
Subsequently, we employed the ResNet-50 model to classify  \num[round-precision=0]{5000} generated images, manually revising the labels to create a new training dataset $\text{D}_2$, which consists of \num[round-precision=0]{1000} realistic (including  \num[round-precision=0]{400} experimental) and \num[round-precision=0]{1000} fake images per pattern. As discussed in \cref{sec: enssemble classification}, this approach significantly improved classification accuracy and precision compared to step 1 against the (same) validation dataset, demonstrating that a larger, curated training dataset enhances model performance. 
Steps 1 and 2 took approximately \num[round-precision=0]{8} hours with our labeling infrastructure. 
If needed, step 2 can be repeated to establish an even larger training dataset.

\paragraph{Step 3: enhancing classification results using ensemble classification strategies}

Since different deep learning models can capture features of different resolutions and characteristics, it might further improve the overall classification performance by combining the predictions from an assortment of models. 
For example, convolutional neural network (CNN) models naturally capture neighborhood locality at different scales (such as VGG) and inter-relationship among various feature extractions (such as ResNet) \cite{he2015deepresiduallearningimage, simonyan2015deepconvolutionalnetworkslargescale, Goodfellow-et-al-2016}.
Whereas, Vision Transformer (ViT) models consider the global spatial relationships of receptive fields (patches) because of the attention blocks \cite{dosovitskiy2021imageworth16x16words}.
In this work, we re-trained six CNN models (ImageNet weights), i.e., ResNet-50, AlexNet, VGG, SqueezeNet, DenseNet, and Inception, and two pre-trained ViT models of different resolutions, namely, \emph{google/vit-base-patch16-224-in21k} and \emph{google/vit-base-patch32-224-in21k}.
We explored a hard voting strategy and two soft voting strategies to combine predictions from all classifiers. 
In the hard voting, the predictions for each class label are summed, and the class with the majority of votes across all classifiers is chosen as the final prediction. 
Conversely, in soft voting, the predicted probabilities for each class label are summed, and the final prediction is determined by the class with the highest weighted probability among all classifiers. 
This can be done using equal weights (average soft voting) or unequal weights (weighted soft voting).%
The individual and ensemble classification results are discussed in \cref{fig: classification results}. 

\subsection{Implementation details}
\paragraph{Fine-tuning foundational stable diffusion model} Our Diffusers model was trained using two NVIDIA RTX A5000 GPUs with Accelerate. 
The images are grayscale with default input dimensions of \(512\times512\) pixels. 
Due to memory limitations, we set the batch size to \num[round-precision=0]{8} and trained the model for \num[round-precision=0]{200} epochs, with the entire training process completing in approximately 1 hour.

\paragraph{Training computer vision models with human annotations} The batch size was set to \num[round-precision=0]{32}, with \SI[round-precision=0]{20}{\percent} of the data reserved for validation. Pre-trained weights from ImageNet were used, and the Adam optimizer with a learning rate of \num[round-precision=3]{0.001} was employed. Categorical cross-entropy was used as the loss function.

\section{Results and discussion}
\subsection{Images generated from the stable diffusion model}
\label{sec: generated images}

\begin{figure*}[h!]
\centering
\begin{tabular}{m{.5em}c@{\hskip 0.12in}c}
& \textcolor{Green}{generated realistic} & \textcolor{Red}{generated fake}
\\ 
\multirow{2}*{\rotatebox{90}{rings}}
&
\includegraphics[width=0.15\linewidth]{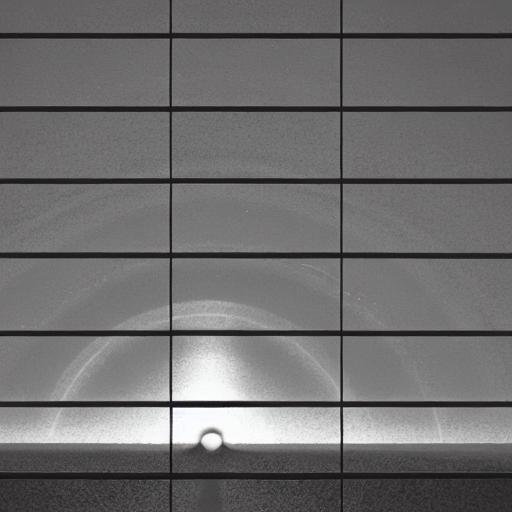}
\includegraphics[width=0.15\linewidth]{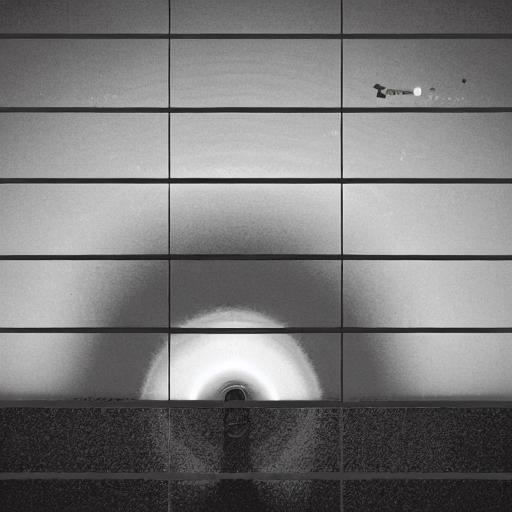}
\includegraphics[width=0.15\linewidth]{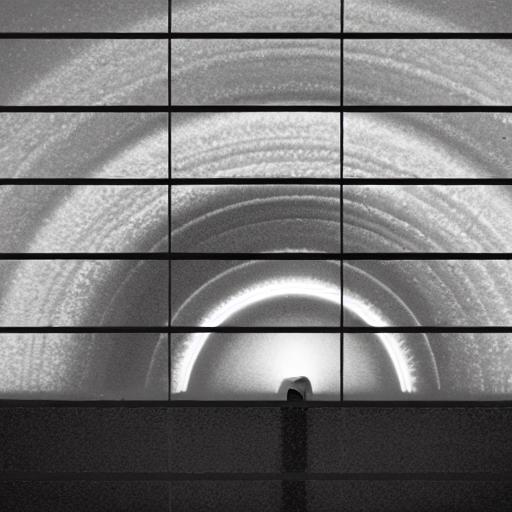}
&
\includegraphics[width=0.15\linewidth]{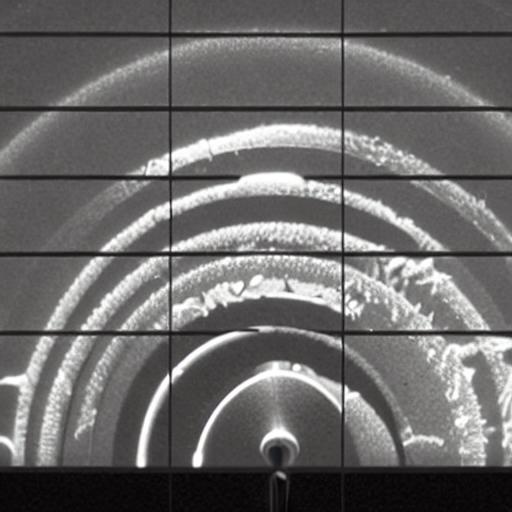}
\includegraphics[width=0.15\linewidth]{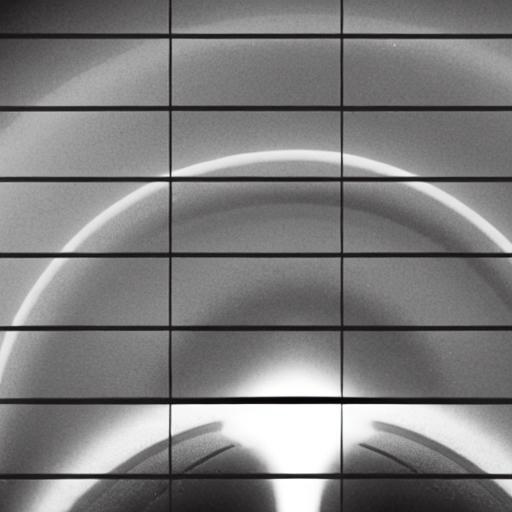}
\includegraphics[width=0.15\linewidth]{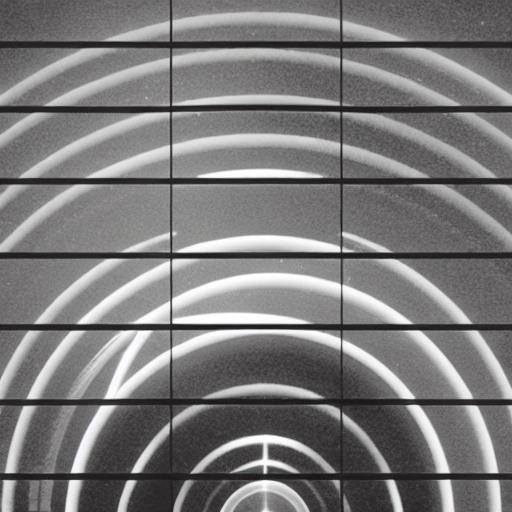}
\\
&
\includegraphics[width=0.15\linewidth]{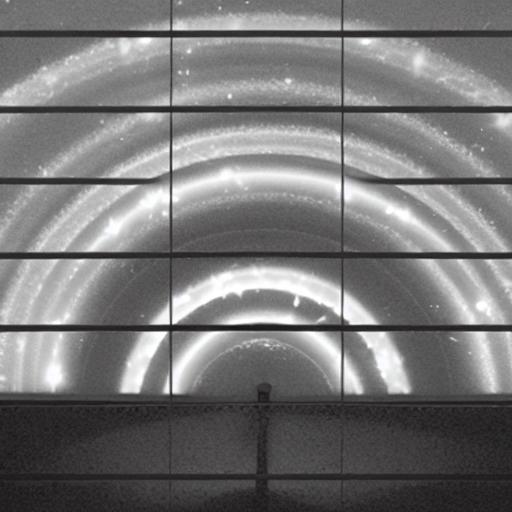}
\includegraphics[width=0.15\linewidth]{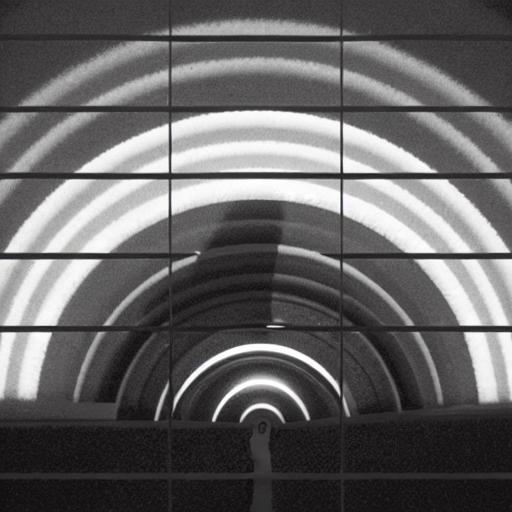}
\includegraphics[width=0.15\linewidth]{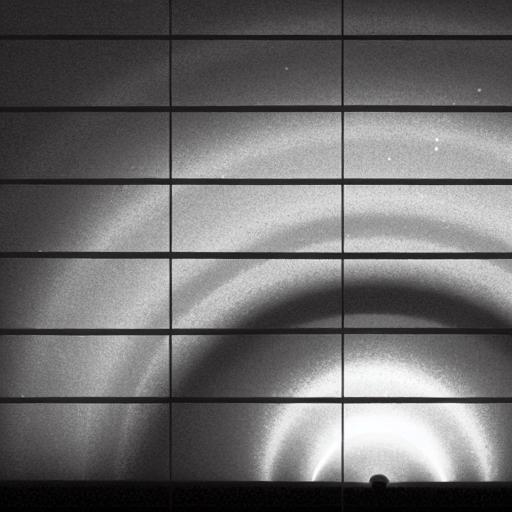}
&
\includegraphics[width=0.15\linewidth]{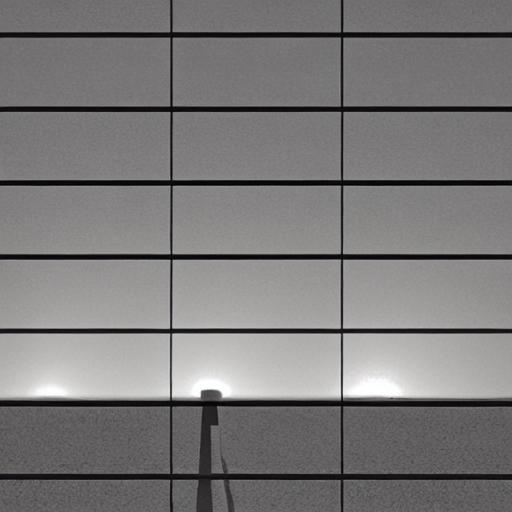}
\includegraphics[width=0.15\linewidth]{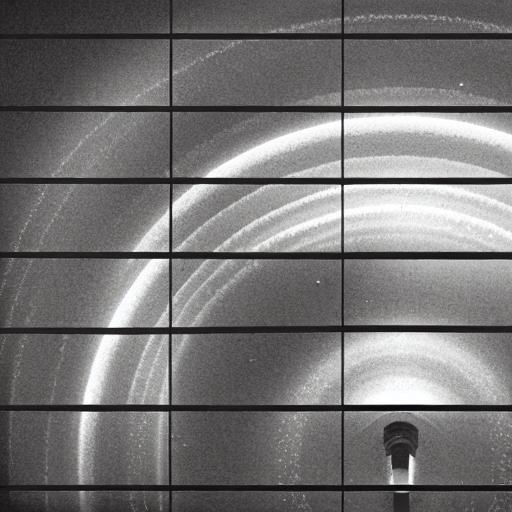}
\includegraphics[width=0.15\linewidth]{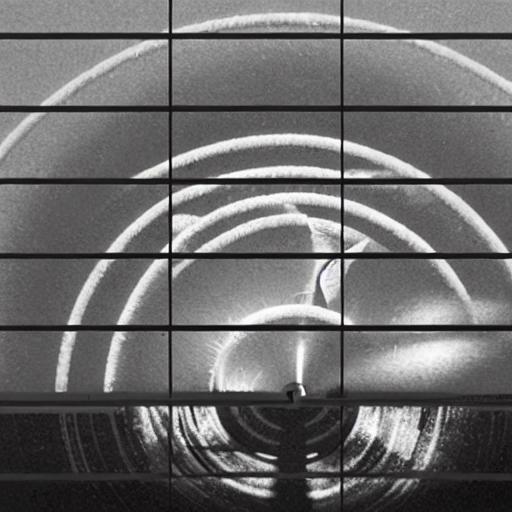}
\\\\
\multirow{2}*{\rotatebox{90}{peaks}}
&
\includegraphics[width=0.15\linewidth]{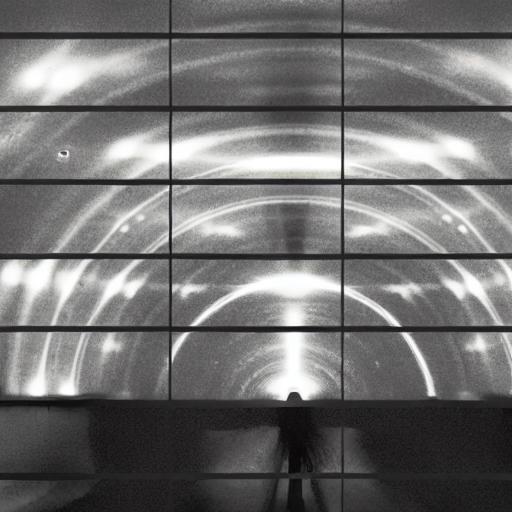}
\includegraphics[width=0.15\linewidth]{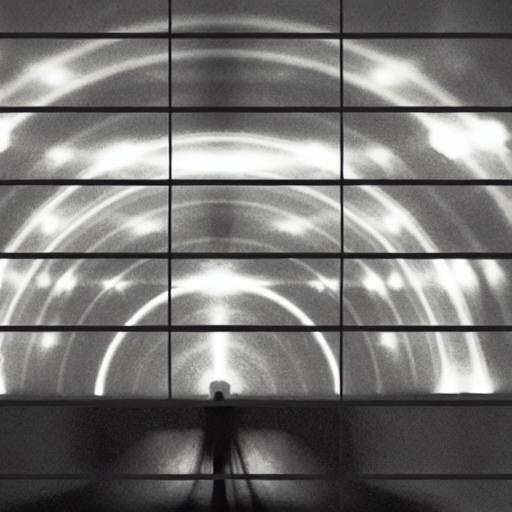}
\includegraphics[width=0.15\linewidth]{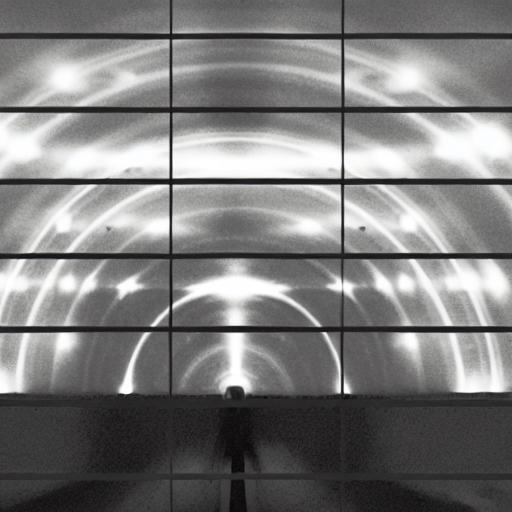}
&
\includegraphics[width=0.15\linewidth]{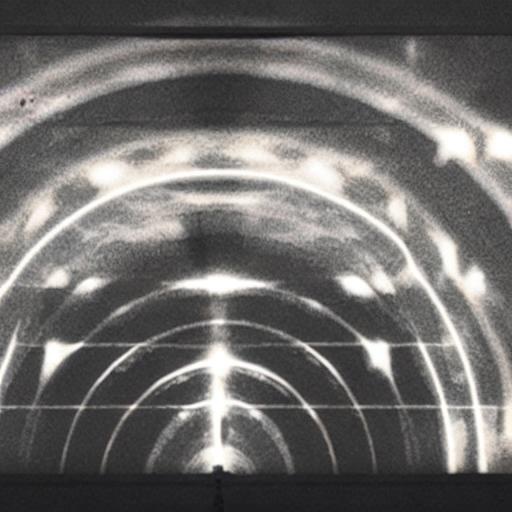}
\includegraphics[width=0.15\linewidth]{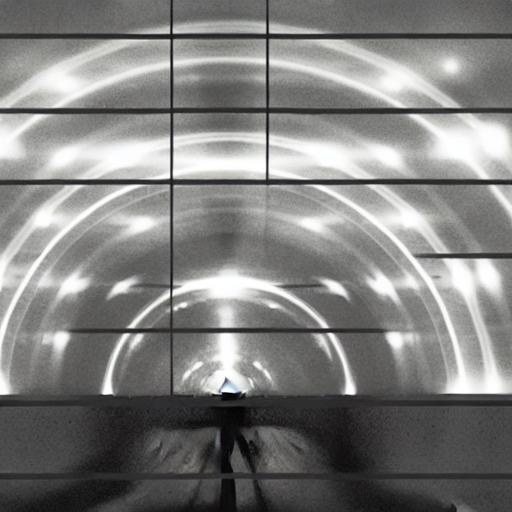}
\includegraphics[width=0.15\linewidth]{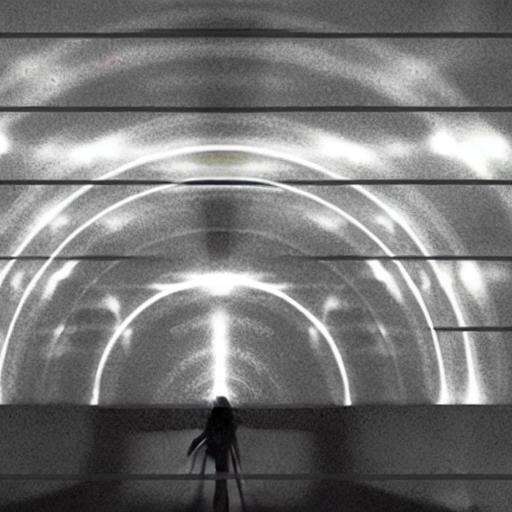}
\\
&
\includegraphics[width=0.15\linewidth]{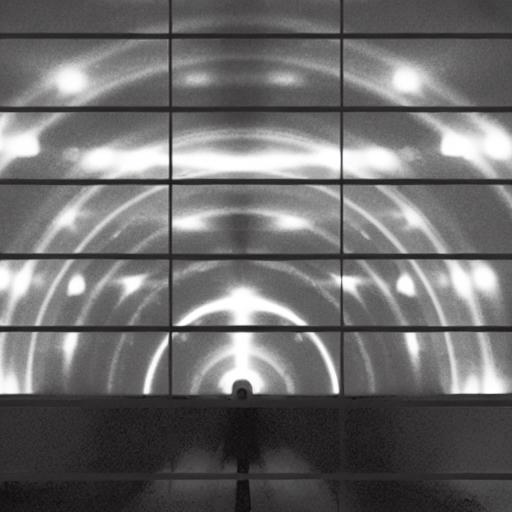}
\includegraphics[width=0.15\linewidth]{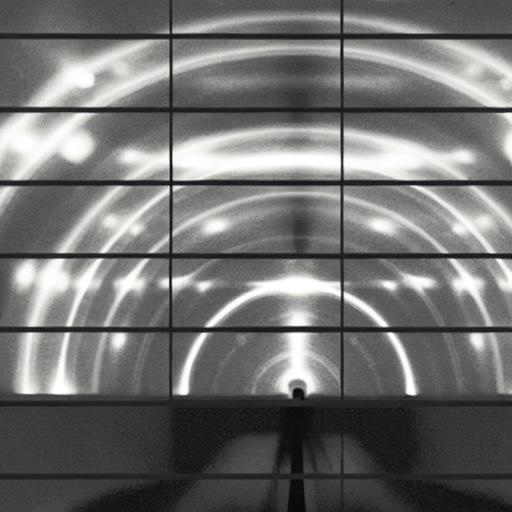}
\includegraphics[width=0.15\linewidth]{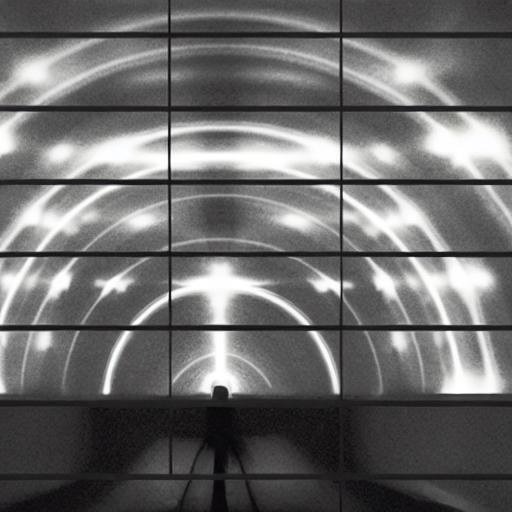}
&
\includegraphics[width=0.15\linewidth]{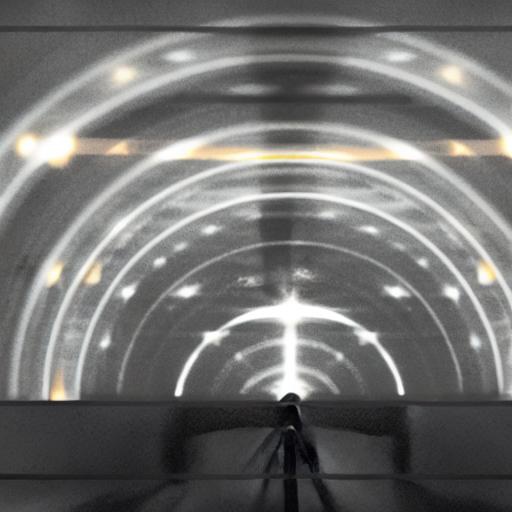}
\includegraphics[width=0.15\linewidth]{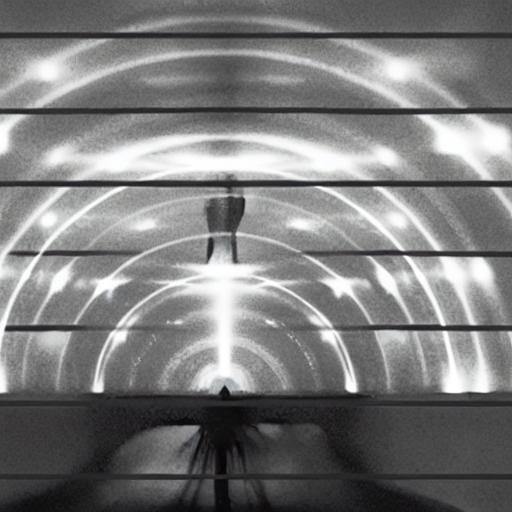}
\includegraphics[width=0.15\linewidth]{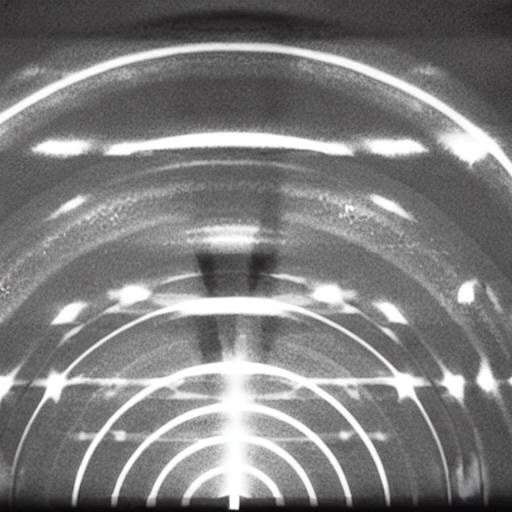}
\\\\
\multirow{2}*{\rotatebox{90}{background}}
&
\includegraphics[width=0.15\linewidth]{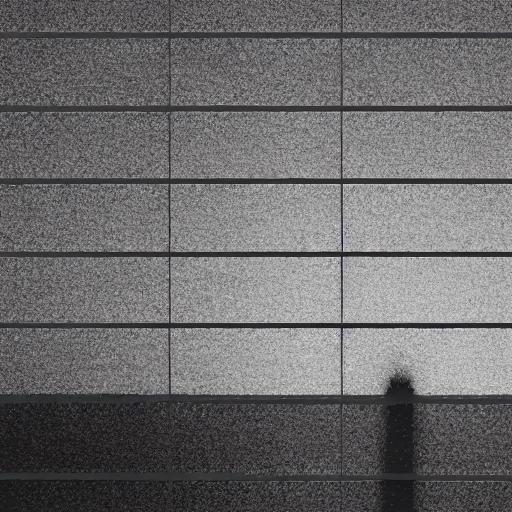}
\includegraphics[width=0.15\linewidth]{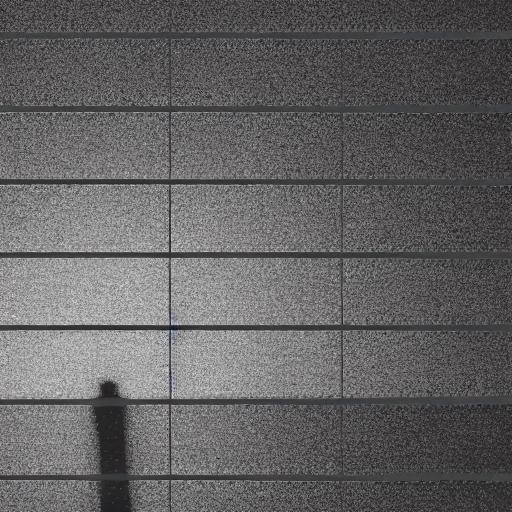}
\includegraphics[width=0.15\linewidth]{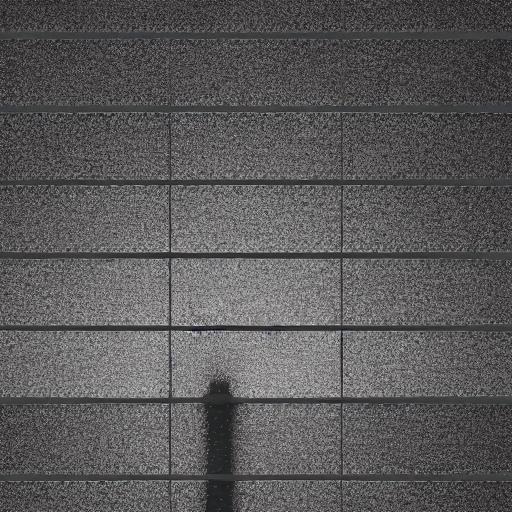}
&
\includegraphics[width=0.15\linewidth]{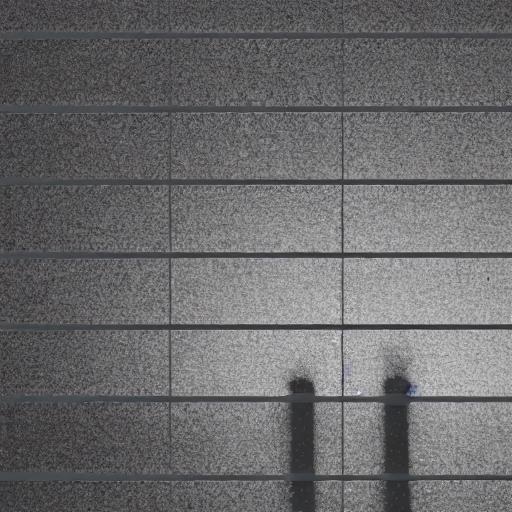}
\includegraphics[width=0.15\linewidth]{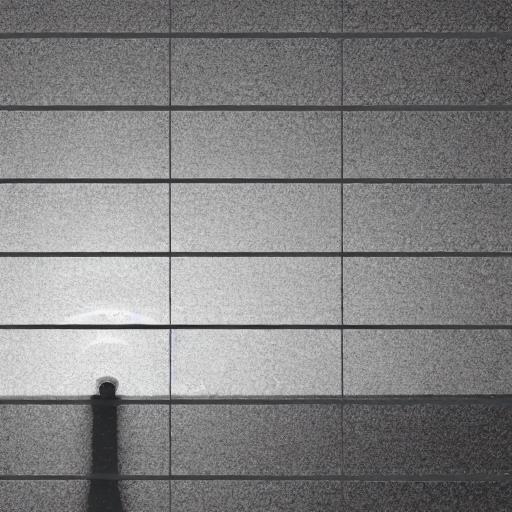}
\includegraphics[width=0.15\linewidth]{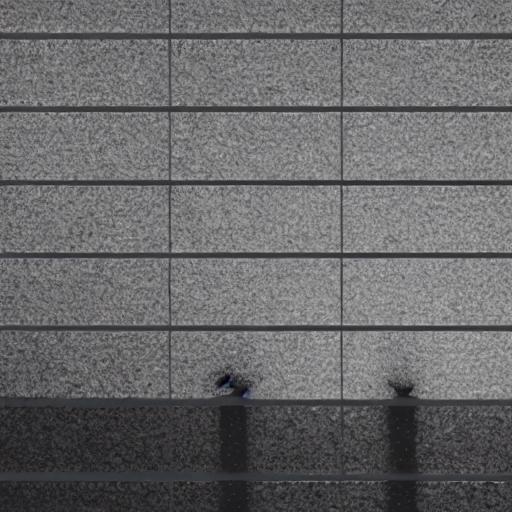}
\\
&
\includegraphics[width=0.15\linewidth]{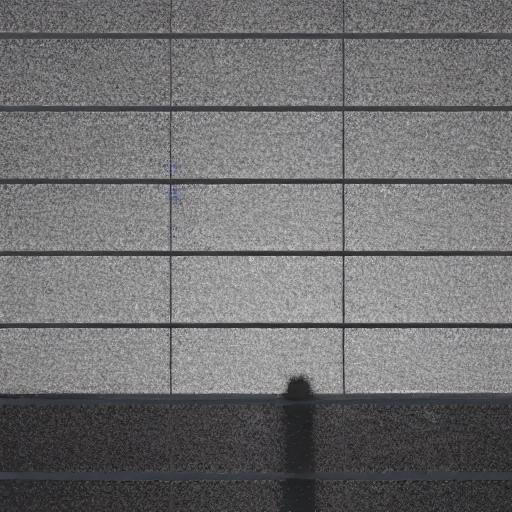}
\includegraphics[width=0.15\linewidth]{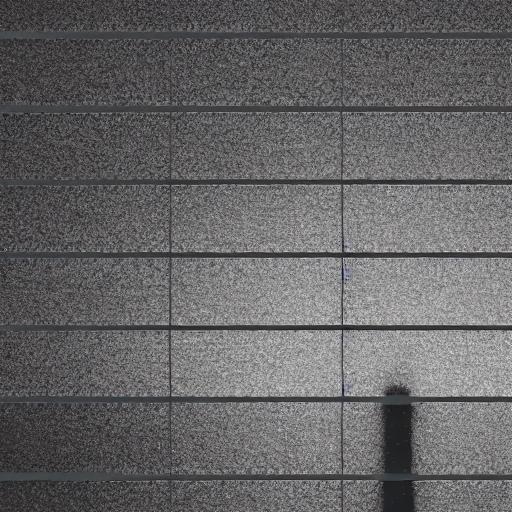}
\includegraphics[width=0.15\linewidth]{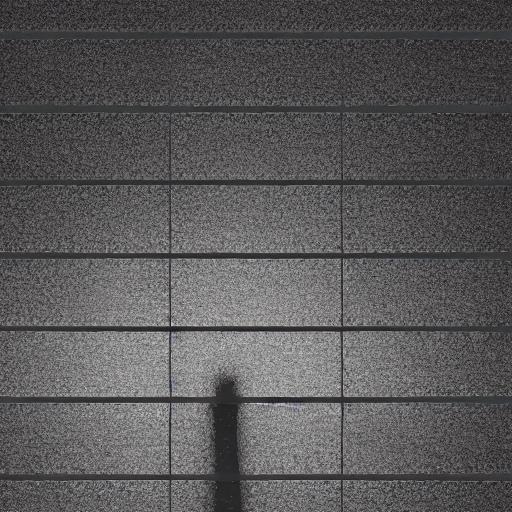}
&
\includegraphics[width=0.15\linewidth]{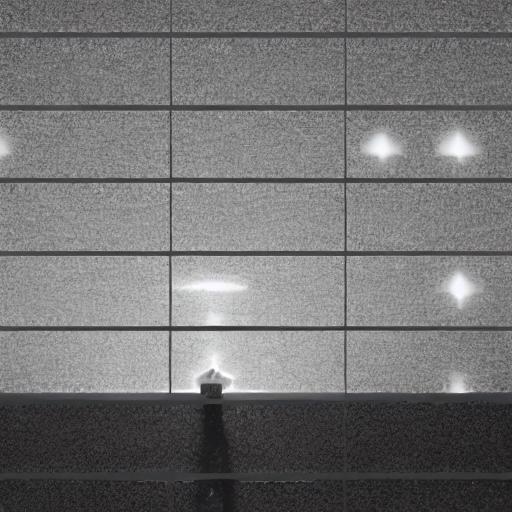}
\includegraphics[width=0.15\linewidth]{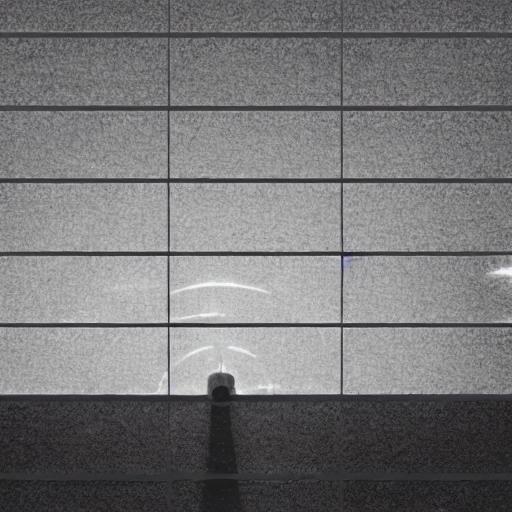}
\includegraphics[width=0.15\linewidth]{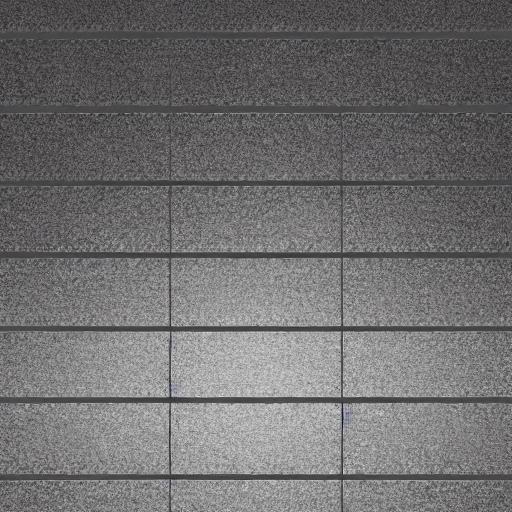}
\end{tabular}

\caption{The realistic and non-realistic (fake) X-ray scattering patterns generated by the fine-tuned Diffusers model.}
\label{fig: generated images}
\end{figure*}

\num[round-precision=0]{40000} images were generated for each pattern using the fine-tuned Diffusers model. 
\Cref{fig: generated images} illustrates that the generated images display both realistic and fake characteristics across the three patterns. 
The following metrics were used to assess the realisticity of the images:

\begin{itemize}
	\setlength\itemsep{-0.2em}
    \item Symmetry: realistic scattering features should exhibit good symmetry with respect to the beam stop.
    \item Continuity: realistic rings should demonstrate good continuity without apparent interruptions or mis-alignments. 
    \item Detector gaps: detector gaps should be straight lines. Missing gaps are \emph{acceptable} since they can be added.
    \item Missing features: minor missing features are \emph{acceptable} because they can be filled using in-painting methods \cite{Chavez:jl5040}.
\end{itemize}

\begin{table}[h!]
\captionof{table}{
Fr\'echet Inception Distance (FID), Kernel Inception Distance (KID), and Inception Score (IS) for the generated images of the three patterns
}
\label{tbl: realisticity metrics}
\begin{center}
\renewcommand{\arraystretch}{1.25}
\begin{tabularx}{.9\linewidth}{l *{5}{>{\centering\arraybackslash}X}}
  \toprule
   \textbf{pattern}
   &
   \textbf{FID}
   & \multicolumn{2}{c}{\textbf{KID}}
   & \multicolumn{2}{c}{\textbf{IS}} 
    \\
  \cmidrule{3-6} 
  & & mean & std & mean & std  \\
  \midrule
  rings &\num[round-precision = 4]{0.6237} & \num[round-precision = 4]{0.0007} & \num[round-precision = 4]{0.0005} & \num[round-precision = 4]{1.9673} & \num[round-precision = 4]{0.1862} \\
  peaks & \num[round-precision = 4]{0.9613} & \num[round-precision = 4]{0.0019} & \num[round-precision = 4]{0.0004} & \num[round-precision = 4]{1.4866}  & \num[round-precision = 4]{0.0918} \\
  background &\num[round-precision = 4]{-8.0062e-8} &\num[round-precision = 4]{-2.5954e-5} & \num[round-precision = 4]{0.0001} & \num[round-precision = 4]{1.5090} & \num[round-precision = 4]{0.1081} \\
\bottomrule
\end{tabularx}
\end{center}
\end{table}

Accessing the realisticity of the generated images has been a challenging topic in the computer vision community \cite{xu2018empiricalstudyevaluationmetrics, lu2023seeingbelievingbenchmarkinghuman, Buzzaccarini_etal2024, podell2023sdxl}. 
There are several metrics commonly used for evaluating the realisticity of the generative models.
For example, both Fr\'echet Inception Distance (FID) \cite{heusel2018ganstrainedtimescaleupdate} and Kernel Inception Distance (KID) \cite{binkowski2021demystifyingmmdgans} use Inception V3 to measure the dissimilarity between generated and real images, with lower scores indicating stronger similarity.
In contrast, Inception Score (IS) does not rely on the statistics of real-world samples but compares the statistics of the generated images, with higher values indicating stronger similarity \cite{salimans2016improvedtechniquestraininggans}.
As shown in \cref{tbl: realisticity metrics}, these scores consistently suggest that the generated background and peaks images are significantly more realistic than rings.
Furthermore, their standard deviation is also smaller compared with rings, which is likely due to the selected training dataset for rings has higher diversity compared to peaks and background patterns (as discussed in \cref{sec: training datasets}). 
Therefore, the generated peaks and background images are more homogeneous and exhibit relatively greater realisticity than the rings.

\begin{figure*}[h!]
\centering
\begin{tabular}{>{\centering\arraybackslash}m{0.01\linewidth} >{\centering\arraybackslash}m{0.2\linewidth} >{\centering\arraybackslash}m{0.2\linewidth} >{\centering\arraybackslash}m{0.2\linewidth} >{\centering\arraybackslash}m{0.2\linewidth}}
& experimental rings & generated rings & experimental peaks & generated peaks \\
\rotatebox{90}{scattering image} &
\includegraphics[width=\linewidth]{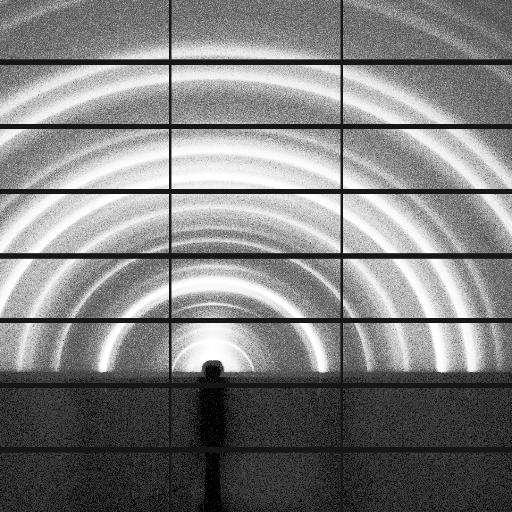} &
\includegraphics[width=\linewidth]{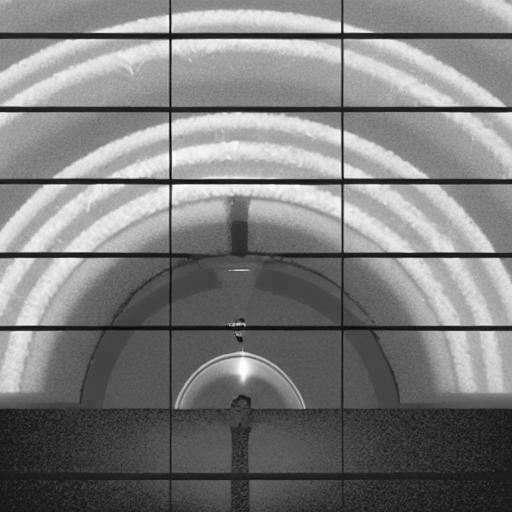} &
\includegraphics[width=\linewidth]{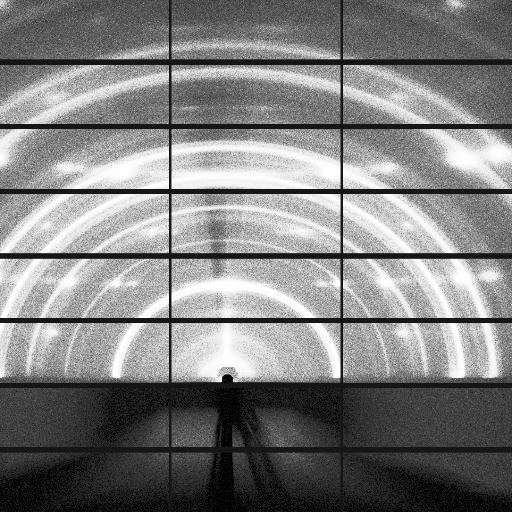} &
\includegraphics[width=\linewidth]{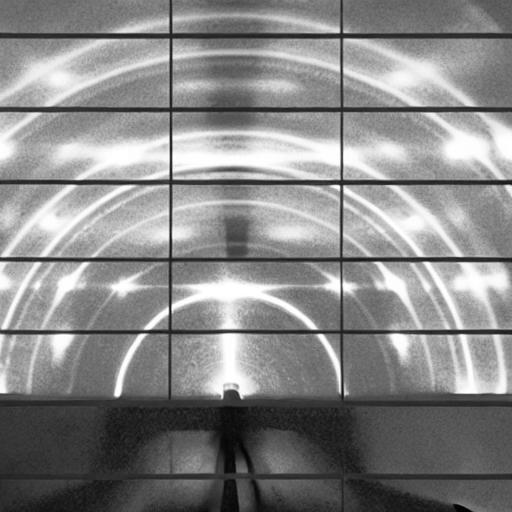} \\\\
\rotatebox{90}{warp transformation} &
\includegraphics[width=\linewidth]{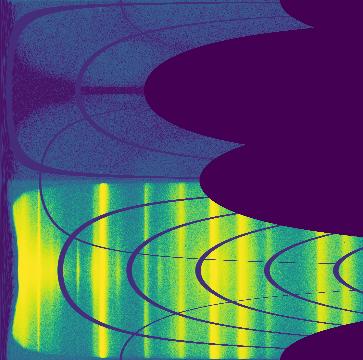} &
\includegraphics[width=\linewidth]{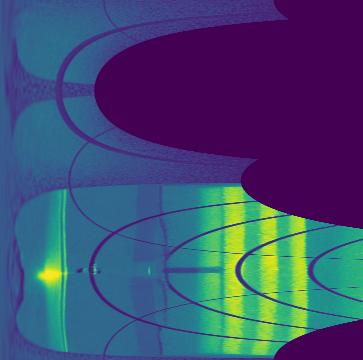} &
\includegraphics[width=\linewidth]{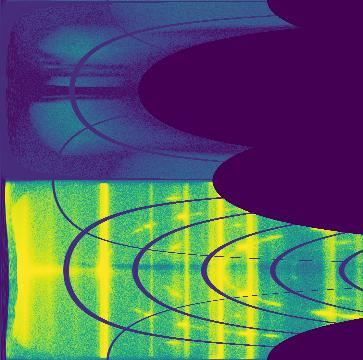} &
\includegraphics[width=\linewidth]{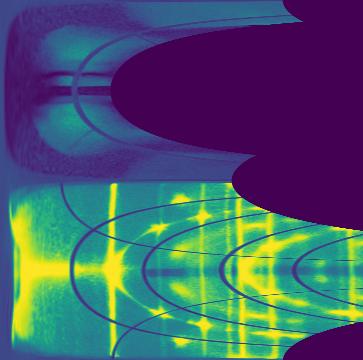} \\

\end{tabular}
\caption{Warp transformation to polar coordinates applied to the selected experimental and generated images}
\label{fig:warp}
\end{figure*}

In addition to the metrics commonly-used for general internet images, we evaluated the realisticity of X-ray scattering images by verifying whether the generated patterns follow the diffraction law.
When an X-ray beam interacts with atoms in a material, it causes the X-rays to be diffracted and scattered at specific angles determined by the atomic spacing and arrangement.
The diffracted X-rays form a series of cones with their apex at the sample position. 
When these cones intersect with the detector, they appear as concentric circles. 
The positions and intensities of these circular patterns should provide information about the atomic structure of the crystalline material.
We transformed these circular patterns from Cartesian coordinates to polar coordinates, which should result in straight lines. 
Examples of these transformation results are presented in \cref{fig:warp}.%
\footnote{The pattern center point for each image was identified using the grid search method, and the inverse transformation was performed using the \emph{warp\_polar} method from the \emph{scikit-image} library \cite{scikit-image}.} 
The vertical appearance of the lines in the transformed images indicates that the generated images exhibit a symmetry that conforms to the diffraction law to some degree.

\subsection{Classification results}
\label{fig: classification results}

\cref{tbl: classification results} presents the values of four metrics, i.e., accuracy, precision, recall, and F1 score, used to assess the classification performance of the eight computer vision models and three voting methods.
Additionally, we evaluated the classification results from the training sets $\text{D}_1$ (1st round) and $\text{D}_2$ (2nd round) in two training rounds, where $\text{D}_2$ is ten times the size of $\text{D}_1$ and was established through the continuous labeling framework described in \cref{sec: enssemble classification}.
In the 2nd round training, the models were trained with $\text{D}_2$ both from scratch and by continuing from the checkpoints trained with $\text{D}_1$.

\begin{align*}
&\vn{Accuracy}  = \frac{\vn{True\,Positives\,(TP)} + \vn{True\,Negtives\,(TN)}}{\vn{TP}+\vn{TN}+\vn{False\,Positives\,(FP)}+\vn{False\,Negatives\,(FN)}}\\[\jot]
&\vn{Precision} = \frac{\vn{TP}}{\vn{TP}+\vn{FP}}\; ,\quad \vn{Recall} = \frac{\vn{TP}}{\vn{TP}+\vn{FN}}\; ,\quad
\vn{F_1} = 2\,\frac{\vn{Precision}\cdot\vn{Recall}}%
                       {\vn{Precision}+\vn{Recall}}
\label{eq: classification metrics}
\end{align*}

Several observations can be made from \cref{tbl: classification results}.
First, Vision Transformers consistently rank the highest among individual models across all metrics.
Second, training with $\text{D}_2$ yields significantly better results compared with $\text{D}_1$, indicating that the proposed continuous training framework does enhance classification performance.
In addition, we noticed that training models with $\text{D}_1$ from scratch leads to better results than continuing training from the 1st round's checkpoints, further demonstrating the importance of establishing high-quality training data.

To leverage the diverse perspectives of different models, we combined their predictions as discussed in \cref{sec: enssemble classification}.
Given the low cost of generating images from a well-trained Diffusers model, achieving high True Positives (TP) and low False Positives (FP) are the primary considerations for this task. 
Consequently, a high precision score is the most desired objective.
Our findings indicate that soft voting with continuous rounds is the most effective strategy among those explored.


\begin{table}[h!]
\captionof{table}{
Classification evaluation of generated ring patterns using eight computer vision models and three voting methods, all trained with labeled datasets utilizing the proposed continuous human labeling framework.
}
\label{tbl: classification results}
\begin{center}
\renewcommand{\arraystretch}{1.3}
\begin{tabularx}{1\linewidth}{lc *{12}{>{\centering\arraybackslash}X}}
  \toprule
   \textbf{classifier}
   &
   \textbf{epochs}
   & \multicolumn{4}{c}{\textbf{1st round}}
   & \multicolumn{4}{c}{\textbf{2nd round (scratch)}} 
   & \multicolumn{4}{c}{\textbf{2nd round (finetune)}} 
    \\
  \cmidrule{3-14} 
  & & acc & prec & F1 & recall & acc & prec & F1 & recall & acc & prec & F1 & recall \\
  \midrule
  ViT-16x16 & 1000 &\num{0.78} & \num{0.7916} & \num{0.7755} & \num{0.76} & \num{0.85} & \num{0.8645} &\num{0.8469} & \num{0.83} & \num{0.835} & \num{0.8383} & \num{0.8341} & \num{0.83} \\
  ViT-32x32 & 1000  &\num{0.79} & \num{0.8020} & \num{0.7857} & \num{0.77} & \num{0.865} & \num{0.8613} &\num{0.8656} & \num{0.87} & \num{0.85} & \num{0.8431} & \num{0.8514} & \textbf{\num{0.86}} \\
  ResNet-50 & 100  &\num{0.755} & \num{0.7575} & \num{0.7537} & \num{0.75} & \num{0.775} & \num{0.8571} &\num{0.7457} & \num{0.66} & \num{0.795} & \num{0.8641} & \num{0.7773} & \num{0.7} \\
  AlexNet & 100 &\num{0.74} & \num{0.7790} & \num{0.7204} & \num{0.67} & \num{0.8} & \num{0.8191} &\num{0.7938} & \num{0.77} & \num{0.79} & \num{0.8085} & \num{0.7835} & \num{0.76} \\
  VGG-11 & 100 &\num{0.775} & \num{0.7894} & \num{0.7692} & \num{0.75} & \num{0.8} & \num{0.7941} &\num{0.8019} & \num{0.81} & \num{0.81} & \num{0.7719} & \num{0.8224} & \num{0.88} \\
  SqueezeNet & 100 &\num{0.75} & \num{0.7604} & \num{0.7448} & \num{0.73} & \num{0.84} & \num{0.8269} &\num{0.8431} & \num{0.86} & \num{0.825} & \num{0.7981} & \num{0.8325} & \num{0.87} \\
  DenseNet-121 & 100 &\num{0.775} & \textbf{\num{0.8767}} & \num{0.7398} & \num{0.64} & \num{0.805} & \num{0.8351} &\num{0.7958} & \num{0.76} & \num{0.84} & \num{0.8541} & \num{0.8367} & \num{0.82} \\
  Inception V3 & 100 &\num{0.705} & \num{0.7157} & \num{0.6974} & \num{0.68} & \num{0.78} & \num{0.7641} &\num{0.7860} & \num{0.81} & \num{0.785} & \num{0.7938} & \num{0.7817} & \num{0.77} \\
  Hard voting & & \num{0.795} & \num{0.8390} & \num{0.7807} & \num{0.73} & \num{0.84} &\num{0.8469} & \num{0.8383} & \num{0.83} & \num{0.825} & \num{0.8350} & \num{0.8223} & \num{0.81} \\
  Soft voting (average) & &\num{0.80} & \num{0.8260} & \num{0.7916} & \num{0.76} & \num{0.84} & \num{0.8469} &\num{0.8383} & \num{0.83} & \num{0.83} & \num{0.8173} & \num{0.8333} & \num{0.85} \\
  Soft voting (weighted) & & \textbf{\num{0.830}} & \num{0.8510} & \textbf{\num{0.8247}} & \textbf{\num{0.80}} & \textbf{\num{0.88}} & \textbf{\num{0.8877}} &\num{0.8787} & \num{0.87} & \num{0.86} & \num{0.8673} & \num{0.8585} & \num{0.85} \\
\bottomrule
\end{tabularx}
\end{center}
\end{table}

\subsection{Visualizing image distribution in latent space}
\begin{figure}
    \centering
    \includegraphics[width=0.15\linewidth]{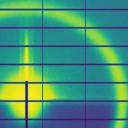}
    \includegraphics[width=0.15\linewidth]{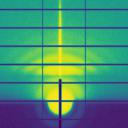}
    \includegraphics[width=0.15\linewidth]{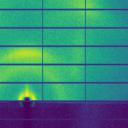}
    \includegraphics[width=0.15\linewidth]{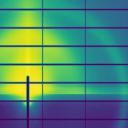}
    \includegraphics[width=0.15\linewidth]{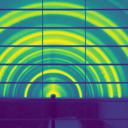}
    \includegraphics[width=0.15\linewidth]{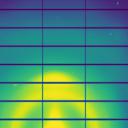} 
    \\
    \includegraphics[width=0.15\linewidth]{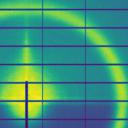}
    \includegraphics[width=0.15\linewidth]{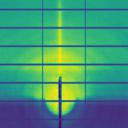}
    \includegraphics[width=0.15\linewidth]{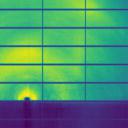}
    \includegraphics[width=0.15\linewidth]{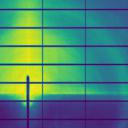}
    \includegraphics[width=0.15\linewidth]{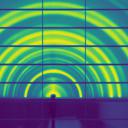}
    \includegraphics[width=0.15\linewidth]{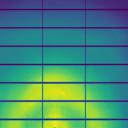} 
    \\\bigskip\bigskip
    \includegraphics[width=0.93\linewidth]{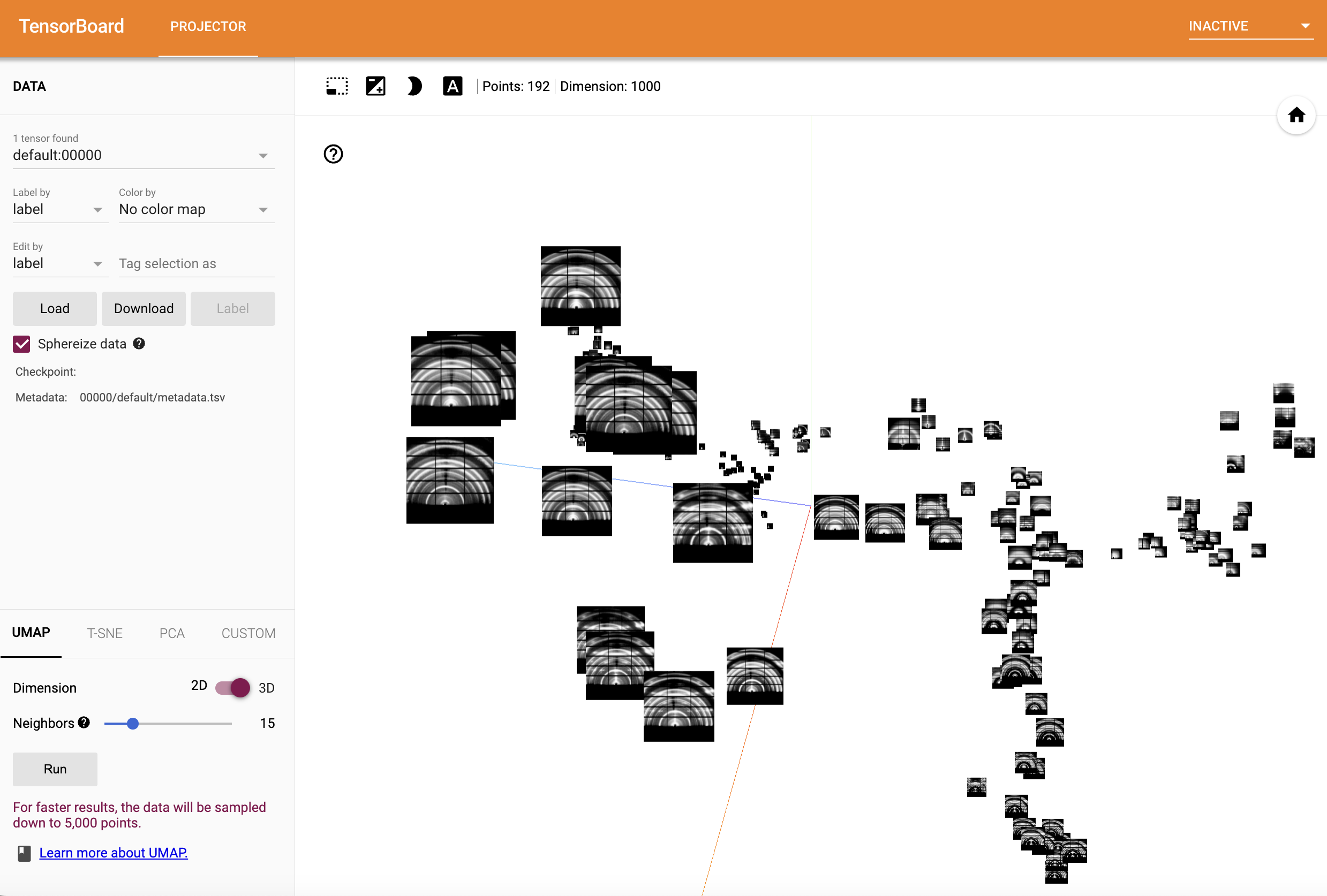}
    \caption{Top row: examples of experimental input images for training a CNN autoencoder. Middle row: the corresponding reconstructed images from the trained autoencoder. Bottom row: 3D UMAP projection of X-ray scattering images in the embedding (latent vector) space.}
    \label{fig: exp image latent visualization}
\end{figure}

The Diffusers model used in this work is a latent diffusion model. 
To visualize the latent vector distribution of experimental images, the selected training dataset, and the generated images (both realistic and fake) across the three patterns (background, rings, and peaks), we projected their latent vectors using UMAP.  
The steps to create the latent vector projections are as follows:

\paragraph{(a) Training an image encoder:} we adopted a self-supervised training strategy and used \num[round-precision=0]{1000} experimental images to train a CNN autoencoder, aiming to minimize the deviation between the input and reconstructed images. 
As shown in \cref{fig: exp image latent visualization}, the reconstructed images (middle row) closely match the original input images (top row), indicating successful training of the autoencoder. 
The encoder was then used to transform each image into a latent vector \(z \in \mathbb{R}^{1000}\).

\paragraph{(b) UMAP projection:} UMAP is essentially a projection of a weighted graph.
The graph is constructed by growing ``seeds'' to form clusters based on the ``connectivity'' between data points in high dimensional space (nearest neighbor descent algorithm) \cite{mcinnes2020umapuniformmanifoldapproximation}.
It is important to note that due to the stochastic nature of UMAP graph construction, the latent vector distributions may vary across different projections. However, the relative relationships among the data points are preserved. 
The bottom row in \cref{fig: exp image latent visualization} displays the 3D projection of latent vectors using UMAP, overlaid with the original images. 
It is apparent that images with similar patterns tend to cluster together.

\begin{figure}[h!]
\hspace*{-0.5cm} 
\centering
\begin{tabular}{m{0.1\linewidth}@{\quad}m{0.28\linewidth}@{\quad}m{0.28\linewidth}@{\quad}m{0.28\linewidth}}
    \vtop{\subfigimg[width=.9\linewidth]{}{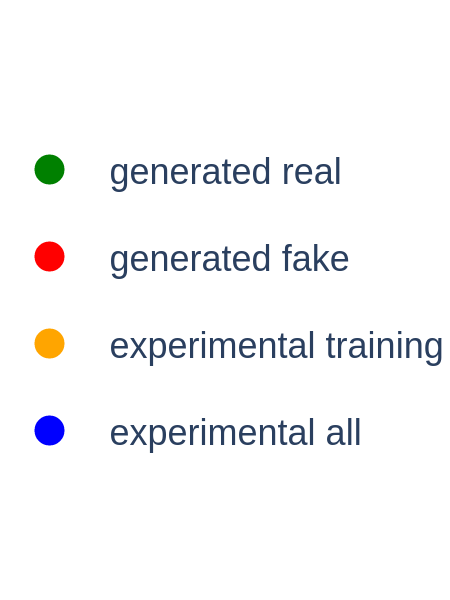}{}} &
    \vtop{\subfigimg[width=\linewidth]{(a)}{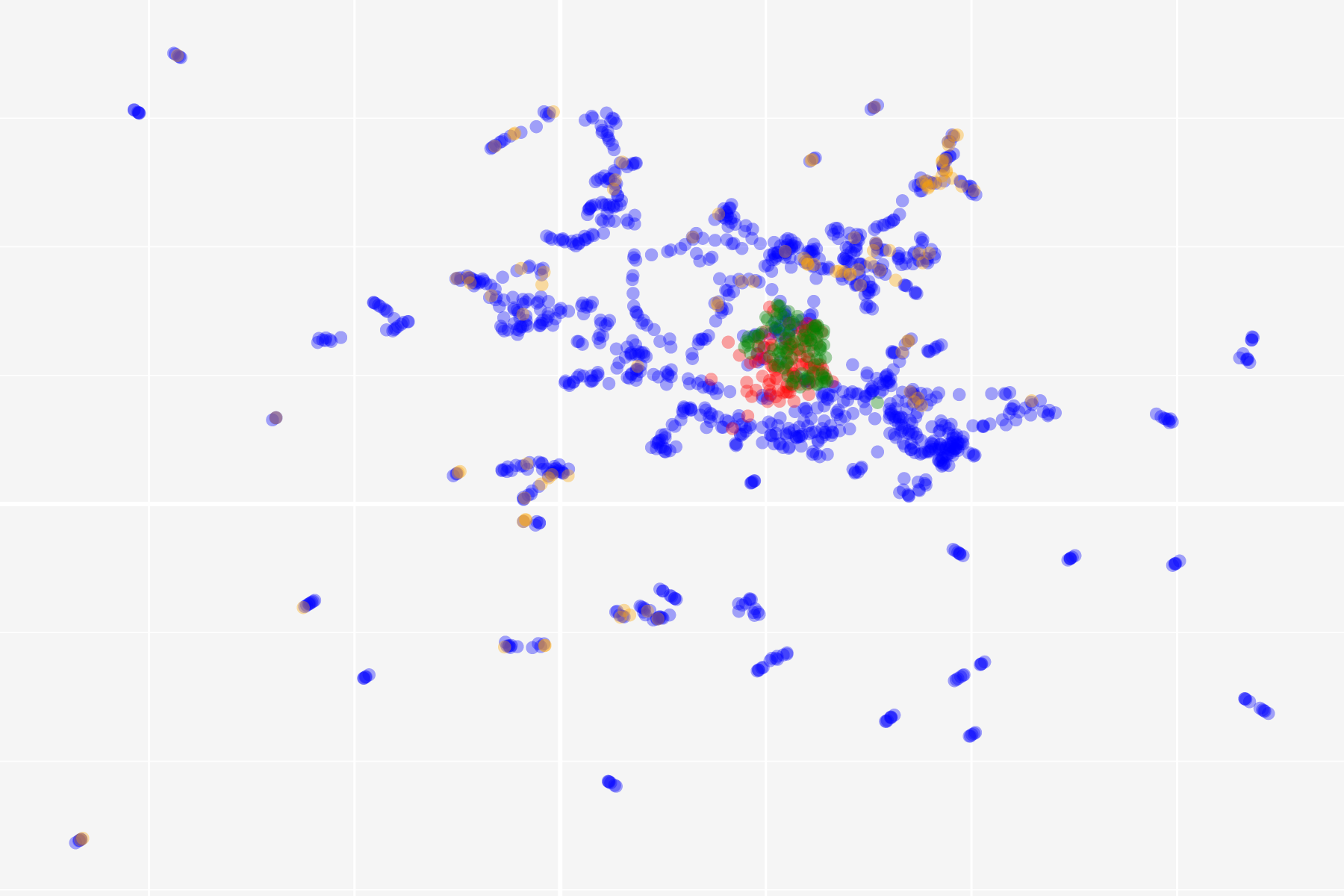}{generated rings}} &
    \vtop{\subfigimg[width=\linewidth]{(b)}{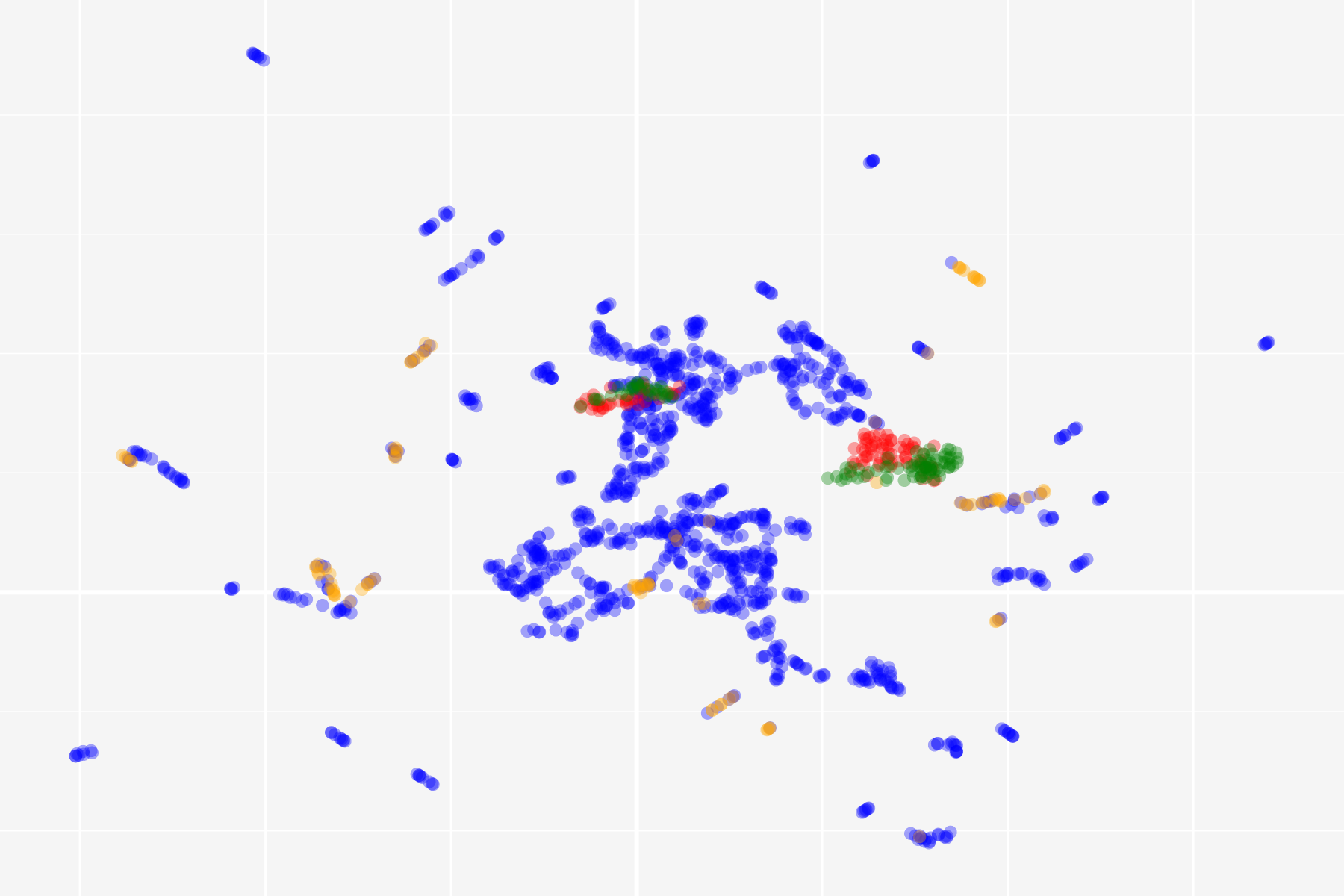}{generated peaks}} &
    \vtop{\subfigimg[width=\linewidth]{(c)}{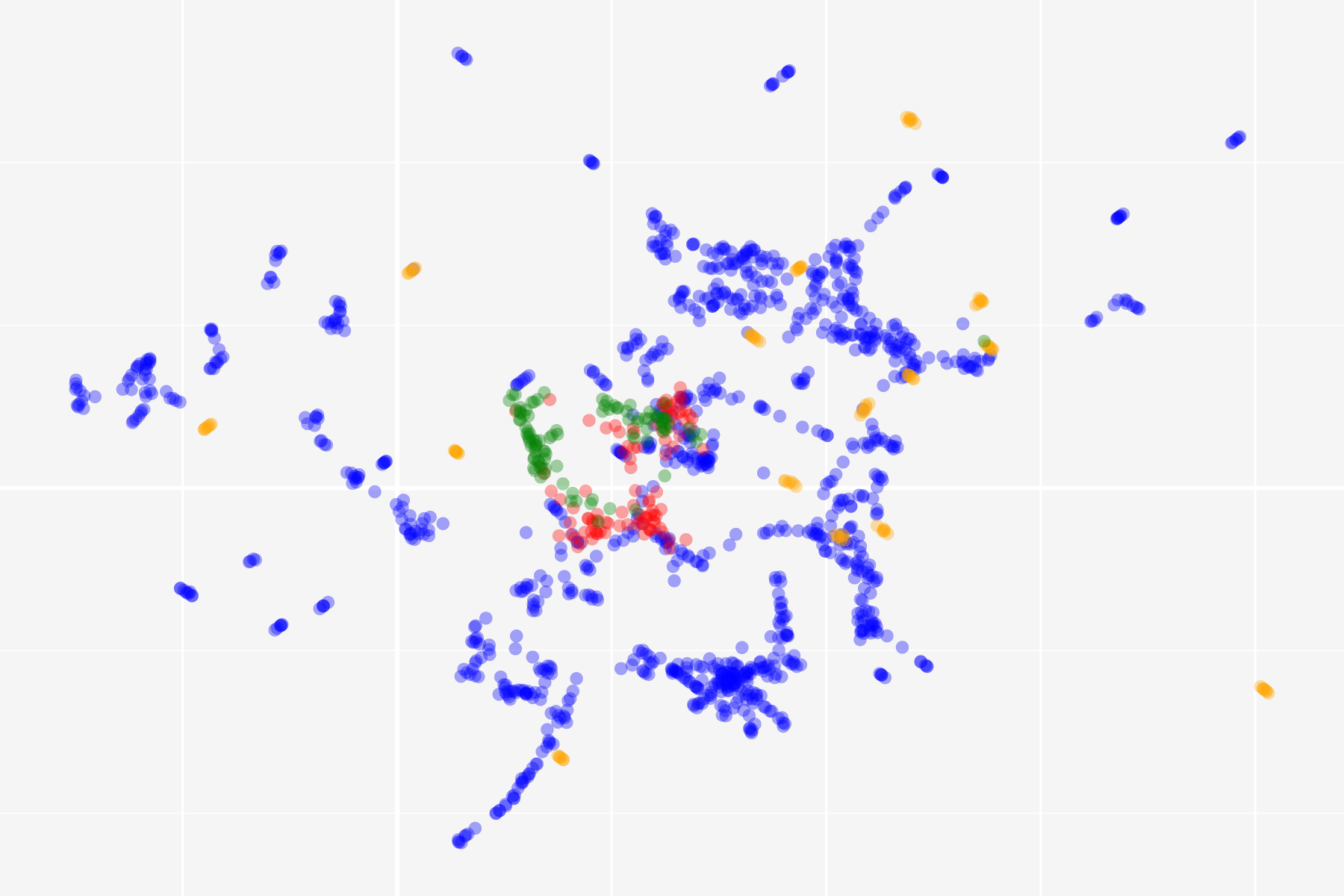}{generated background}} \\\\
    \vtop{\subfigimg[width=\linewidth]{}{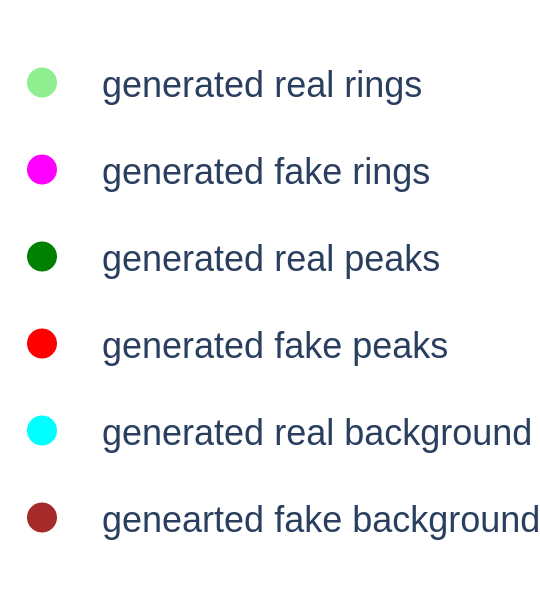}{}} &
    \multicolumn{3}{@{}m{0.89\linewidth}@{}}{\subfigimg[width=\linewidth]{(d)}{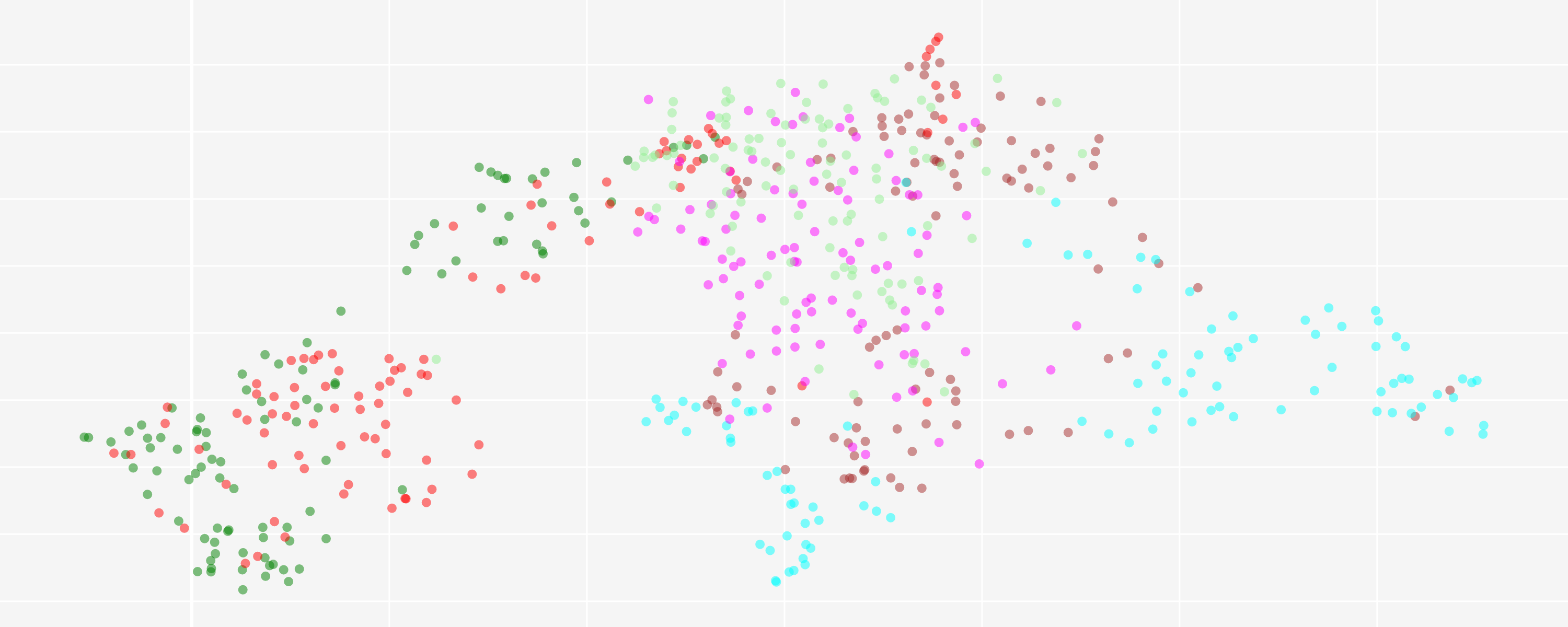}{all three generated patterns}} \\
\end{tabular}
\caption{2D UMAP projection of latent vectors of X-ray scattering images for different generated patterns. Note that all experiment data (blue) and experimental training data (orange) in a-c are the same datasets resulting from different projections.}
\label{fig: generated image latent visualization}
\end{figure}

\Cref{fig: generated image latent visualization}a-c display the 2D projections of the latent vectors of the entire experimental dataset (blue, randomly sampled \num[round-precision=0]{1000} out of \num[round-precision=0]{400000} images) and the training dataset (orange, \num[round-precision=0]{100} images); the generated images by the Diffusers model are shown as realistic (green) and fake (red) for rings, peaks, and background, respectively.
Several observations can be made.

For rings (\cref{fig: generated image latent visualization}a), despite our training images being fairly evenly distributed compared to all experimental images, the generated images (both red and green) cluster in a much smaller region near the center.
This indicates that the diffusion model tends to generalize the training dataset distribution, reducing inhomogeneity among the generated images.
This observation relates to an open challenge in generative AI---increasing the diversity of text-image models \cite{Kumari_2023_CVPR}.

For peaks (\cref{fig: generated image latent visualization}b), the training images primarily scatter in the top right and bottom left regions. 
Consistent with this, the generated peaks images also cluster in these two regions in the same directions.

Interestingly, the training images for the background (\cref{fig: generated image latent visualization}c) are distributed in the outer space of the entire experimental dataset, whereas the generated images still cluster near the center. 
For all three patterns, the realistic and fake generated images have overlaps, highlighting the challenge in classifying the generated images.

\Cref{fig: generated image latent visualization}d shows the distribution of the generated images of all three patterns in the same 2D UMAP projection. 
The peaks form the most separable clusters, whereas rings and background images exhibit more overlap, though they remain generally separable. 
This demonstrates that the trained diffusion model is capable of generalizing distinct features of different patterns.

\subsection{Aspects of improvement}
To further improve the realisticity of generated images, one strategy is to refine the prompts and generate prompt alternatives using generative language models. 
This ensures two key outcomes: (a) the generated images align more closely with the intended context and requirements, resulting in higher-quality outputs tailored to specific scientific applications; (b) more descriptive text-image datasets can be used to integrate the foundational diffusion model into continuous training cycles, ultimately leading to a better diffusion-classifier framework capable of accommodating more general descriptions.
Another potential enhancement is to leverage more advanced diffusion models, such as Diffusers XL.
Additionally, training a smaller diffusion model and text encoder from scratch exclusively for a specific domain might be a viable approach.

\section{Conclusions}
In this study, we established a text-to-image pipeline to generate X-ray scattering images using a fine-tuned foundational stable diffusion model. 
To enhance the realisticity of the generated images, we trained an ensemble of computer vision classifiers to identify the most realistic images. 
Several implementation details contributed to the improved results:
\begin{enumerate}
    \item We trained the classifiers iteratively, incorporating annotators to interactively annotate and correct the classified images.
    \item The training set for each iteration was composed of a mixture of experimental and generated images, with a ratio of 4 to 6.
    \item The optimal results were achieved by adopting a weighted soft voting strategy that combines predictions from an ensemble of computer vision models.
\end{enumerate}

The pipeline demonstrated the efficacy and affordability of using generative AI for synthesizing scientific images. 
It is anticipated that this technique will be highly beneficial for scientific applications where data scarcity is a limiting factor. 
For instance, generative AI can be valuable in situations requiring extensive datasets, such as training and refining domain-specific foundational models. 
Moreover, it holds great potential for enhancing educational experiences, e.g., providing realistic training environments for remote facility users.

While generative AI has been explored broadly to facilitate scientific discovery in areas including reasoning and fetching useful knowledge \cite{LIU2023798} and making scientific plans \cite{huang2024mlagentbenchevaluatinglanguageagents}, we recognize it can pose significant risks to the scientific community, potentially compromising scientific safety and integrity.
We align with the five principles of human accountability and responsibility for scientific efforts employing AI \cite{Blau_etal2024}. 
AI-generated image data should not replace experimental data in scientific analyses, and we advocate for the adoption of more regulations to harness this powerful tool responsibly.

\section*{Code availability}
The code used in this study is available on GitHub at \url{https://github.com/mlexchange/mlex_scientific_txt2image}.

\section*{Acknowledgements}
This work is supported by U.S. Department of Energy, Office of Science, Office of Basic Energy Sciences Data, Artificial Intelligence, and Machine Learning at DOE Scientific User Facilities program under the MLExchange Project (Award Number 107514).

\bibliographystyle{IEEEtran}
\bibliography{paper}

\begin{thebibliography}{10}
\providecommand{\url}[1]{#1}
\csname url@samestyle\endcsname
\providecommand{\newblock}{\relax}
\providecommand{\bibinfo}[2]{#2}
\providecommand{\BIBentrySTDinterwordspacing}{\spaceskip=0pt\relax}
\providecommand{\BIBentryALTinterwordstretchfactor}{4}
\providecommand{\BIBentryALTinterwordspacing}{\spaceskip=\fontdimen2\font plus
\BIBentryALTinterwordstretchfactor\fontdimen3\font minus
  \fontdimen4\font\relax}
\providecommand{\BIBforeignlanguage}[2]{{%
\expandafter\ifx\csname l@#1\endcsname\relax
\typeout{** WARNING: IEEEtran.bst: No hyphenation pattern has been}%
\typeout{** loaded for the language `#1'. Using the pattern for}%
\typeout{** the default language instead.}%
\else
\language=\csname l@#1\endcsname
\fi
#2}}
\providecommand{\BIBdecl}{\relax}
\BIBdecl

\bibitem{CORREABAENA20181410}
\BIBentryALTinterwordspacing
J.-P. Correa-Baena, K.~Hippalgaonkar, J.~{van Duren}, S.~Jaffer, V.~R.
  Chandrasekhar, V.~Stevanovic, C.~Wadia, S.~Guha, and T.~Buonassisi,
  ``Accelerating materials development via automation, machine learning, and
  high-performance computing,'' \emph{Joule}, vol.~2, no.~8, pp. 1410--1420,
  2018. [Online]. Available:
  \url{https://www.sciencedirect.com/science/article/pii/S2542435118302289}
\BIBentrySTDinterwordspacing

\bibitem{mlexchange2022}
\BIBentryALTinterwordspacing
Z.~Zhao, T.~Chavez, E.~A. Holman, G.~Hao, A.~Green, H.~Krishnan, D.~McReynolds,
  R.~J. Pandolfi, E.~J. Roberts, P.~H. Zwart, H.~Yanxon, N.~Schwarz,
  S.~Sankaranarayanan, S.~V. Kalinin, A.~Mehta, S.~I. Campbell, and A.~Hexemer,
  ``\uppercase{MLE}xchange: A web-based platform enabling exchangeable machine
  learning workflows for scientific studies,'' in \emph{2022 4th Annual
  Workshop on Extreme-scale Experiment-in-the-Loop Computing (XLOOP)}, 2022,
  pp. 10--15. [Online]. Available: \url{10.1109/XLOOP56614.2022.00007}
\BIBentrySTDinterwordspacing

\bibitem{autonomous_microscope2023}
\BIBentryALTinterwordspacing
S.~V. Kalinin, D.~Mukherjee, K.~Roccapriore, B.~J. Blaiszik, A.~Ghosh, M.~A.
  Ziatdinov, A.~Al-Najjar, C.~Doty, S.~Akers, N.~S. Rao, J.~C. Agar, and S.~R.
  Spurgeon, ``Machine learning for automated experimentation in scanning
  transmission electron microscopy,'' \emph{npj Computational Materials},
  vol.~9, p. 227, 2023. [Online]. Available:
  \url{https://doi.org/10.1038/s41524-023-01142-0}
\BIBentrySTDinterwordspacing

\bibitem{ZHAO202071}
\BIBentryALTinterwordspacing
Z.~Zhao, T.~R. Bieler, J.~LLorca, and P.~Eisenlohr, ``Grain boundary slip
  transfer classification and metric selection with artificial neural
  networks,'' \emph{Scripta Materialia}, vol. 185, pp. 71--75, 2020. [Online].
  Available:
  \url{https://www.sciencedirect.com/science/article/pii/S1359646220302499}
\BIBentrySTDinterwordspacing

\bibitem{Zubatiuk_etal2021}
\BIBentryALTinterwordspacing
T.~Zubatiuk and O.~Isayev, ``Development of multimodal machine learning
  potentials: Toward a physics-aware artificial intelligence,'' \emph{Accounts
  of Chemical Research}, vol.~54, no.~7, pp. 1575--1585, 2021, pMID: 33715355.
  [Online]. Available: \url{https://doi.org/10.1021/acs.accounts.0c00868}
\BIBentrySTDinterwordspacing

\bibitem{Amal_etal2022}
\BIBentryALTinterwordspacing
S.~Amal, L.~Safarnejad, J.~A. Omiye, I.~Ghanzouri, J.~H. Cabot, and E.~G. Ross,
  ``Use of multi-modal data and machine learning to improve cardiovascular
  disease care,'' \emph{Frontiers in Cardiovascular Medicine}, vol.~9, 2022.
  [Online]. Available:
  \url{https://www.frontiersin.org/journals/cardiovascular-medicine/articles/10.3389/fcvm.2022.840262}
\BIBentrySTDinterwordspacing

\bibitem{Ektefaie_etal2023}
\BIBentryALTinterwordspacing
Y.~Ektefaie, G.~Dasoulas, A.~Noori, M.~Farhat, and M.~Zitnik, ``Multimodal
  learning with graphs,'' \emph{Nature Machine Intelligence}, vol.~5, no.~4,
  pp. 340--350, 2023. [Online]. Available:
  \url{https://doi.org/10.1038/s42256-023-00624-6}
\BIBentrySTDinterwordspacing

\bibitem{CryoDRGN}
\BIBentryALTinterwordspacing
E.~Zhong, T.~Bepler, B.~Berger, and D.~Davis, ``Cryo\uppercase{DRGN}:
  reconstruction of heterogeneous cryo-\uppercase{EM} structures using neural
  networks,'' \emph{Nature Methods}, vol.~18, no. 176-185, 2021. [Online].
  Available: \url{https://doi.org/10.1038/s41592-020-01049-4}
\BIBentrySTDinterwordspacing

\bibitem{tomotwin}
\BIBentryALTinterwordspacing
G.~Rice, T.~Wagner, and M.~e.~a. Stabrin, ``Tomotwin: generalized 3d
  localization of macromolecules in cryo-electron tomograms with structural
  data mining,'' \emph{Nature Methods}, vol.~20, no. 871–880, 2023. [Online].
  Available: \url{https://doi.org/10.1038/s41592-023-01878-z}
\BIBentrySTDinterwordspacing

\bibitem{bioengineering10121435}
\BIBentryALTinterwordspacing
L.~Pinto-Coelho, ``How artificial intelligence is shaping medical imaging
  technology: A survey of innovations and applications,''
  \emph{Bioengineering}, vol.~10, no.~12, 2023. [Online]. Available:
  \url{https://www.mdpi.com/2306-5354/10/12/1435}
\BIBentrySTDinterwordspacing

\bibitem{tibbers_etal2023}
\BIBentryALTinterwordspacing
G.~Hao, E.~J. Roberts, T.~Chavez, Z.~Zhao, E.~A. Holman, H.~Yanxon, A.~Green,
  H.~Krishnan, D.~Ushizima, D.~McReynolds, N.~Schwarz, P.~H. Zwart, A.~Hexemer,
  and D.~Parkinson, ``Deploying machine learning based segmentation for
  scientific imaging analysis at synchrotron facilities,'' \emph{Electronic
  Imaging}, vol.~35, no.~9, pp. 290--1--290--1, 2023. [Online]. Available:
  \url{https://library.imaging.org/ei/articles/35/9/IPAS-290}
\BIBentrySTDinterwordspacing

\bibitem{khaira2017derivation}
\BIBentryALTinterwordspacing
G.~Khaira, M.~Doxastakis, A.~Bowen, J.~Ren, H.~S. Suh, T.~Segal-Peretz,
  X.~Chen, C.~Zhou, A.~F. Hannon, N.~J. Ferrier \emph{et~al.}, ``Derivation of
  multiple covarying material and process parameters using physics-based
  modeling of \uppercase{X}-ray data,'' \emph{Macromolecules}, vol.~50, no.~19,
  pp. 7783--7793, 2017. [Online]. Available:
  \url{https://pubs.acs.org/doi/10.1021/acs.macromol.7b00691}
\BIBentrySTDinterwordspacing

\bibitem{guo2022physics}
\BIBentryALTinterwordspacing
Z.~Guo, J.~K. Song, G.~Barbastathis, M.~E. Glinsky, C.~T. Vaughan, K.~W.
  Larson, B.~K. Alpert, and Z.~H. Levine, ``Physics-assisted generative
  adversarial network for \uppercase{X}-ray tomography,'' \emph{Opt. Express},
  vol.~30, no.~13, pp. 23\,238--23\,259, Jun 2022. [Online]. Available:
  \url{https://opg.optica.org/oe/abstract.cfm?URI=oe-30-13-23238}
\BIBentrySTDinterwordspacing

\bibitem{pmlr-v80-mescheder18a}
\BIBentryALTinterwordspacing
L.~Mescheder, A.~Geiger, and S.~Nowozin, ``Which training methods for {GAN}s do
  actually converge?'' in \emph{Proceedings of the 35th International
  Conference on Machine Learning}, ser. Proceedings of Machine Learning
  Research, J.~Dy and A.~Krause, Eds., vol.~80.\hskip 1em plus 0.5em minus
  0.4em\relax PMLR, 10--15 Jul 2018, pp. 3481--3490. [Online]. Available:
  \url{https://proceedings.mlr.press/v80/mescheder18a.html}
\BIBentrySTDinterwordspacing

\bibitem{10565846}
M.~Welfert, G.~R. Kurri, K.~Otstot, and L.~Sankar, ``Addressing gan training
  instabilities via tunable classification losses,'' \emph{IEEE Journal on
  Selected Areas in Information Theory}, vol.~5, pp. 534--553, 2024.

\bibitem{NEURIPS2020_3eb46aa5}
\BIBentryALTinterwordspacing
Y.~Wu, P.~Zhou, A.~G. Wilson, E.~Xing, and Z.~Hu, ``Improving gan training with
  probability ratio clipping and sample reweighting,'' in \emph{Advances in
  Neural Information Processing Systems}, H.~Larochelle, M.~Ranzato,
  R.~Hadsell, M.~Balcan, and H.~Lin, Eds., vol.~33.\hskip 1em plus 0.5em minus
  0.4em\relax Curran Associates, Inc., 2020, pp. 5729--5740. [Online].
  Available:
  \url{https://proceedings.neurips.cc/paper_files/paper/2020/file/3eb46aa5d93b7a5939616af91addfa88-Paper.pdf}
\BIBentrySTDinterwordspacing

\bibitem{bai2022traininghelpfulharmlessassistant}
\BIBentryALTinterwordspacing
Y.~Bai, A.~Jones, K.~Ndousse, A.~Askell, A.~Chen, N.~DasSarma, D.~Drain,
  S.~Fort, D.~Ganguli, T.~Henighan, N.~Joseph, S.~Kadavath, J.~Kernion,
  T.~Conerly, S.~El-Showk, N.~Elhage, Z.~Hatfield-Dodds, D.~Hernandez, T.~Hume,
  S.~Johnston, S.~Kravec, L.~Lovitt, N.~Nanda, C.~Olsson, D.~Amodei, T.~Brown,
  J.~Clark, S.~McCandlish, C.~Olah, B.~Mann, and J.~Kaplan, ``Training a
  helpful and harmless assistant with reinforcement learning from human
  feedback,'' 2022. [Online]. Available: \url{https://arxiv.org/abs/2204.05862}
\BIBentrySTDinterwordspacing

\bibitem{brown2020languagemodelsfewshotlearners}
\BIBentryALTinterwordspacing
T.~B. Brown, B.~Mann, N.~Ryder, M.~Subbiah, J.~Kaplan, P.~Dhariwal,
  A.~Neelakantan, P.~Shyam, G.~Sastry, A.~Askell, S.~Agarwal, A.~Herbert-Voss,
  G.~Krueger, T.~Henighan, R.~Child, A.~Ramesh, D.~M. Ziegler, J.~Wu,
  C.~Winter, C.~Hesse, M.~Chen, E.~Sigler, M.~Litwin, S.~Gray, B.~Chess,
  J.~Clark, C.~Berner, S.~McCandlish, A.~Radford, I.~Sutskever, and D.~Amodei,
  ``Language models are few-shot learners,'' 2020. [Online]. Available:
  \url{https://arxiv.org/abs/2005.14165}
\BIBentrySTDinterwordspacing

\bibitem{rombach2022high}
\BIBentryALTinterwordspacing
R.~Rombach, A.~Blattmann, D.~Lorenz, P.~Esser, and B.~Ommer, ``High-resolution
  image synthesis with latent diffusion models,'' in \emph{Proceedings of the
  IEEE/CVF conference on computer vision and pattern recognition}, 2022, pp.
  10\,684--10\,695. [Online]. Available:
  \url{https://www.computer.org/csdl/proceedings-article/cvpr/2022/694600k0674/1H1iFsO7Zuw}
\BIBentrySTDinterwordspacing

\bibitem{liu2024sorareviewbackgroundtechnology}
\BIBentryALTinterwordspacing
Y.~Liu, K.~Zhang, Y.~Li, Z.~Yan, C.~Gao, R.~Chen, Z.~Yuan, Y.~Huang, H.~Sun,
  J.~Gao, L.~He, and L.~Sun, ``Sora: A review on background, technology,
  limitations, and opportunities of large vision models,'' 2024. [Online].
  Available: \url{https://arxiv.org/abs/2402.17177}
\BIBentrySTDinterwordspacing

\bibitem{maaz2024videochatgptdetailedvideounderstanding}
\BIBentryALTinterwordspacing
M.~Maaz, H.~Rasheed, S.~Khan, and F.~S. Khan,
  ``Video-\uppercase{C}hat\uppercase{GPT}: Towards detailed video understanding
  via large vision and language models,'' 2024. [Online]. Available:
  \url{https://arxiv.org/abs/2306.05424}
\BIBentrySTDinterwordspacing

\bibitem{ho2020denoisingdiffusionprobabilisticmodels}
\BIBentryALTinterwordspacing
J.~Ho, A.~Jain, and P.~Abbeel, ``Denoising diffusion probabilistic models,''
  2020. [Online]. Available: \url{https://arxiv.org/abs/2006.11239}
\BIBentrySTDinterwordspacing

\bibitem{song2022denoisingdiffusionimplicitmodels}
\BIBentryALTinterwordspacing
J.~Song, C.~Meng, and S.~Ermon, ``Denoising diffusion implicit models,'' 2022.
  [Online]. Available: \url{https://arxiv.org/abs/2010.02502}
\BIBentrySTDinterwordspacing

\bibitem{podell2023sdxl}
D.~Podell, Z.~English, K.~Lacey, A.~Blattmann, T.~Dockhorn, J.~M{\"u}ller,
  J.~Penna, and R.~Rombach, ``\uppercase{Sdxl}: Improving latent diffusion
  models for high-resolution image synthesis,'' \emph{arXiv preprint
  arXiv:2307.01952}, 2023.

\bibitem{videoworldsimulators2024}
\BIBentryALTinterwordspacing
T.~Brooks, B.~Peebles, C.~Holmes, W.~DePue, Y.~Guo, L.~Jing, D.~Schnurr,
  J.~Taylor, T.~Luhman, E.~Luhman, C.~Ng, R.~Wang, and A.~Ramesh, ``Video
  generation models as world simulators,'' 2024. [Online]. Available:
  \url{https://openai.com/research/video-generation-models-as-world-simulators}
\BIBentrySTDinterwordspacing

\bibitem{chambon2022adapting}
\BIBentryALTinterwordspacing
P.~Chambon, C.~Bluethgen, C.~P. Langlotz, and A.~Chaudhari, ``Adapting
  pretrained vision-language foundational models to medical imaging domains,''
  2022. [Online]. Available: \url{https://arxiv.org/abs/2210.04133}
\BIBentrySTDinterwordspacing

\bibitem{hashmi2024xreal}
\BIBentryALTinterwordspacing
A.~U.~R. Hashmi, I.~Almakky, M.~A. Qazi, S.~Sanjeev, V.~R. Papineni,
  D.~Mahapatra, and M.~Yaqub, ``Xreal: Realistic anatomy and pathology-aware
  \uppercase{X}-ray generation via controllable diffusion model,'' 2024.
  [Online]. Available: \url{https://arxiv.org/abs/2403.09240}
\BIBentrySTDinterwordspacing

\bibitem{liang2024covid}
\BIBentryALTinterwordspacing
Z.~Liang, Z.~Xue, S.~Rajaraman, and S.~Antani, ``Covid-19 pneumonia chest x-ray
  pattern synthesis by stable diffusion,'' in \emph{2024 IEEE Southwest
  Symposium on Image Analysis and Interpretation (SSIAI)}.\hskip 1em plus 0.5em
  minus 0.4em\relax Los Alamitos, CA, USA: IEEE Computer Society, mar 2024, pp.
  21--24. [Online]. Available:
  \url{https://doi.ieeecomputersociety.org/10.1109/SSIAI59505.2024.10508671}
\BIBentrySTDinterwordspacing

\bibitem{diffusers}
\BIBentryALTinterwordspacing
Diffusers online documentation. [Online]. Available:
  \url{https://huggingface.co/docs/diffusers/en/index}
\BIBentrySTDinterwordspacing

\bibitem{aithal2024understandinghallucinationsdiffusionmodels}
\BIBentryALTinterwordspacing
S.~K. Aithal, P.~Maini, Z.~C. Lipton, and J.~Z. Kolter, ``Understanding
  hallucinations in diffusion models through mode interpolation,'' 2024.
  [Online]. Available: \url{https://arxiv.org/abs/2406.09358}
\BIBentrySTDinterwordspacing

\bibitem{weng2024hallucination}
\BIBentryALTinterwordspacing
L.~Weng, ``Extrinsic \uppercase{H}allucinations in \uppercase{LLM}s.''
  \emph{lilianweng.github.io}, Jul 2024. [Online]. Available:
  \url{https://lilianweng.github.io/posts/2024-07-07-hallucination/}
\BIBentrySTDinterwordspacing

\bibitem{zhao2022mlexchange}
Z.~Zhao, T.~Chavez, E.~A. Holman, G.~Hao, A.~Green, H.~Krishnan, D.~McReynolds,
  R.~J. Pandolfi, E.~J. Roberts, P.~H. Zwart \emph{et~al.},
  ``\uppercase{MLE}xchange: A web-based platform enabling exchangeable machine
  learning workflows for scientific studies,'' in \emph{2022 4th Annual
  Workshop on Extreme-scale Experiment-in-the-Loop Computing (XLOOP)}.\hskip
  1em plus 0.5em minus 0.4em\relax IEEE, 2022, pp. 10--15.

\bibitem{he2015deepresiduallearningimage}
\BIBentryALTinterwordspacing
K.~He, X.~Zhang, S.~Ren, and J.~Sun, ``Deep residual learning for image
  recognition,'' 2015. [Online]. Available:
  \url{https://arxiv.org/abs/1512.03385}
\BIBentrySTDinterwordspacing

\bibitem{simonyan2015deepconvolutionalnetworkslargescale}
\BIBentryALTinterwordspacing
K.~Simonyan and A.~Zisserman, ``Very deep convolutional networks for
  large-scale image recognition,'' 2015. [Online]. Available:
  \url{https://arxiv.org/abs/1409.1556}
\BIBentrySTDinterwordspacing

\bibitem{Goodfellow-et-al-2016}
I.~Goodfellow, Y.~Bengio, and A.~Courville, \emph{Deep Learning}.\hskip 1em
  plus 0.5em minus 0.4em\relax MIT Press, 2016,
  \url{http://www.deeplearningbook.org}.

\bibitem{dosovitskiy2021imageworth16x16words}
\BIBentryALTinterwordspacing
A.~Dosovitskiy, L.~Beyer, A.~Kolesnikov, D.~Weissenborn, X.~Zhai,
  T.~Unterthiner, M.~Dehghani, M.~Minderer, G.~Heigold, S.~Gelly, J.~Uszkoreit,
  and N.~Houlsby, ``An image is worth 16x16 words: Transformers for image
  recognition at scale,'' 2021. [Online]. Available:
  \url{https://arxiv.org/abs/2010.11929}
\BIBentrySTDinterwordspacing

\bibitem{Chavez:jl5040}
\BIBentryALTinterwordspacing
T.~Chavez, E.~J. Roberts, P.~H. Zwart, and A.~Hexemer, ``{A comparison of
  deep-learning-based inpainting techniques for experimental X-ray
  scattering},'' \emph{Journal of Applied Crystallography}, vol.~55, no.~5, pp.
  1277--1288, Oct 2022. [Online]. Available:
  \url{https://doi.org/10.1107/S1600576722007105}
\BIBentrySTDinterwordspacing

\bibitem{xu2018empiricalstudyevaluationmetrics}
\BIBentryALTinterwordspacing
Q.~Xu, G.~Huang, Y.~Yuan, C.~Guo, Y.~Sun, F.~Wu, and K.~Weinberger, ``An
  empirical study on evaluation metrics of generative adversarial networks,''
  2018. [Online]. Available: \url{https://arxiv.org/abs/1806.07755}
\BIBentrySTDinterwordspacing

\bibitem{lu2023seeingbelievingbenchmarkinghuman}
\BIBentryALTinterwordspacing
Z.~Lu, D.~Huang, L.~Bai, J.~Qu, C.~Wu, X.~Liu, and W.~Ouyang, ``Seeing is not
  always believing: Benchmarking human and model perception of
  \uppercase{AI}-generated images,'' 2023. [Online]. Available:
  \url{https://arxiv.org/abs/2304.13023}
\BIBentrySTDinterwordspacing

\bibitem{Buzzaccarini_etal2024}
\BIBentryALTinterwordspacing
``The promise and pitfalls of \uppercase{AI}-generated anatomical images:
  Evaluating \uppercase{M}idjourney for aesthetic surgery applications,''
  \emph{Aesthetic Plastic Surgery}, vol.~48, no.~9, pp. 1874--1883, 2024.
  [Online]. Available: \url{https://doi.org/10.1007/s00266-023-03826-w}
\BIBentrySTDinterwordspacing

\bibitem{heusel2018ganstrainedtimescaleupdate}
\BIBentryALTinterwordspacing
M.~Heusel, H.~Ramsauer, T.~Unterthiner, B.~Nessler, and S.~Hochreiter,
  ``\uppercase{GAN}s trained by a two time-scale update rule converge to a
  local nash equilibrium,'' 2018. [Online]. Available:
  \url{https://arxiv.org/abs/1706.08500}
\BIBentrySTDinterwordspacing

\bibitem{binkowski2021demystifyingmmdgans}
\BIBentryALTinterwordspacing
M.~Bińkowski, D.~J. Sutherland, M.~Arbel, and A.~Gretton, ``Demystifying
  \uppercase{MMD GAN}s,'' 2021. [Online]. Available:
  \url{https://arxiv.org/abs/1801.01401}
\BIBentrySTDinterwordspacing

\bibitem{salimans2016improvedtechniquestraininggans}
\BIBentryALTinterwordspacing
T.~Salimans, I.~Goodfellow, W.~Zaremba, V.~Cheung, A.~Radford, and X.~Chen,
  ``Improved techniques for training \uppercase{GAN}s,'' 2016. [Online].
  Available: \url{https://arxiv.org/abs/1606.03498}
\BIBentrySTDinterwordspacing

\bibitem{scikit-image}
\BIBentryALTinterwordspacing
S.~van~der Walt, J.~L. Schönberger, J.~Nunez-Iglesias, F.~Boulogne, J.~D.
  Warner, N.~Yager, E.~Gouillart, T.~Yu, and the scikit-image contributors,
  ``scikit-image: image processing in python,'' \emph{PeerJ}, vol.~2, p. e453,
  jun 2014. [Online]. Available: \url{https://doi.org/10.7717/peerj.453}
\BIBentrySTDinterwordspacing

\bibitem{mcinnes2020umapuniformmanifoldapproximation}
\BIBentryALTinterwordspacing
L.~McInnes, J.~Healy, and J.~Melville, ``\uppercase{UMAP}: Uniform manifold
  approximation and projection for dimension reduction,'' 2020. [Online].
  Available: \url{https://arxiv.org/abs/1802.03426}
\BIBentrySTDinterwordspacing

\bibitem{Kumari_2023_CVPR}
N.~Kumari, B.~Zhang, R.~Zhang, E.~Shechtman, and J.-Y. Zhu, ``Multi-concept
  customization of text-to-image diffusion,'' in \emph{Proceedings of the
  IEEE/CVF Conference on Computer Vision and Pattern Recognition (CVPR)}, June
  2023, pp. 1931--1941.

\bibitem{LIU2023798}
\BIBentryALTinterwordspacing
Y.~Liu, Z.~Yang, Z.~Yu, Z.~Liu, D.~Liu, H.~Lin, M.~Li, S.~Ma, M.~Avdeev, and
  S.~Shi, ``Generative artificial intelligence and its applications in
  materials science: Current situation and future perspectives,'' \emph{Journal
  of Materiomics}, vol.~9, no.~4, pp. 798--816, 2023. [Online]. Available:
  \url{https://www.sciencedirect.com/science/article/pii/S2352847823000771}
\BIBentrySTDinterwordspacing

\bibitem{huang2024mlagentbenchevaluatinglanguageagents}
\BIBentryALTinterwordspacing
Q.~Huang, J.~Vora, P.~Liang, and J.~Leskovec, ``Mlagentbench: Evaluating
  language agents on machine learning experimentation,'' 2024. [Online].
  Available: \url{https://arxiv.org/abs/2310.03302}
\BIBentrySTDinterwordspacing

\bibitem{Blau_etal2024}
\BIBentryALTinterwordspacing
W.~Blau, V.~G. Cerf, J.~Enriquez, J.~S. Francisco, U.~Gasser, M.~L. Gray,
  M.~Greaves, B.~J. Grosz, K.~H. Jamieson, G.~H. Haug, J.~L. Hennessy,
  E.~Horvitz, D.~I. Kaiser, A.~J. London, R.~Lovell-Badge, M.~K. McNutt,
  M.~Minow, T.~M. Mitchell, S.~Ness, S.~Parthasarathy, S.~Perlmutter, W.~H.
  Press, J.~M. Wing, and M.~Witherell, ``Protecting scientific integrity in an
  age of generative ai,'' \emph{Proceedings of the National Academy of
  Sciences}, vol. 121, no.~22, p. e2407886121, 2024. [Online]. Available:
  \url{https://www.pnas.org/doi/abs/10.1073/pnas.2407886121}
\BIBentrySTDinterwordspacing

\end{thebibliography}

\end{document}